\documentclass[titlepage,12pt]{article}
\usepackage{amsmath,amssymb,latexsym}

\def\hybrid{
        \topmargin -8pt    
        \oddsidemargin 0pt
        \headheight 0pt
        \headsep 0pt
        \textwidth 6.25in       
        \textheight 9.5in       
        \marginparwidth .875in
        \parskip 5pt plus 1pt   
        \jot = 1.5ex}
\hybrid

\numberwithin{equation}{section}
\numberwithin{table}{section}
\numberwithin{figure}{section}

\DeclareMathOperator{\tr}{tr}
\DeclareMathOperator{\coker}{coker}
\DeclareMathOperator{\Imag}{Im}
\DeclareMathOperator{\Real}{Re}
\newcommand{\id}{\mathbf{1}}
\newcommand{\dd}{\textrm{d}}
\newcommand{\vol}{\textrm{vol}}
\newcommand{\rel}[1]{\underline{#1}}

\newcommand{\abs}[1]{\lvert #1 \rvert}

\newcommand{\bti}[3]{#1_{#2}^{\hphantom{#2}#3}}
\newcommand{\tbundle}[1]{\textrm{T}#1}
\newcommand{\nbundle}[1]{\textrm{N}#1}

\newcommand{\inv}[1]{{#1}^\mathbf{-1}}
\newcommand{\leftup}[2]{\vphantom{#2}^{#1}\!\!{#2}}
\newcommand{\diag}[1]{\text{diag}\left({#1}\right)}
\newcommand{\ac}[2]{\{{#1},{#2}\}}
\newcommand{\super}[1]{\boldsymbol{#1}}
\newcommand{\spinrep}[1]{\boldsymbol{#1}}
\newcommand{\singspin}{\xi}
\newcommand{\sindex}[1]{\mathfrak{#1}}
\newcommand{\spinder}{\mathcal{D}}
\newcommand{\pair}[1]{\vec{#1}}

\newcommand{\db}{\zeta}              
\newcommand{\fcharge}{Q_{\tilde f}}  
\newcommand{\dbbf}{\mathcal{B}}      
\newcommand{\cov}{\nabla}            
\newcommand{\dil}{S}                 

\begin{document}


\begin{titlepage}
\begin{center}

\hfill hep-th/0502059\\ 
\hfill ZMP-HH/05-03\\
\vskip 2.25cm

{\large \bf D-terms and F-terms from D7-brane fluxes}\footnote{%
Work supported by: DFG -- The German Science Foundation,
European RTN Program MRTN-CT-2004-503369 and the
DAAD -- the German Academic Exchange Service.}\\

\vskip 2cm

{\bf Hans Jockers and Jan Louis} \\

\vskip 1.5cm

{\em II. Institut f{\"u}r Theoretische Physik\\
Universit{\"a}t Hamburg\\
Luruper Chaussee 149\\
D-22761 Hamburg, Germany}\\
\vskip 1cm 
{\tt  hans.jockers@desy.de, jan.louis@desy.de} \\
\end{center}

\vskip 2cm

\begin{center} {\bf ABSTRACT } \end{center}
Using a Kaluza-Klein reduction of the fermionic part of the D-brane action we compute D- and F-terms of the $\mathcal{N}=1$ effective action for generic Calabi-Yau orientifold compactifications in the presence of a space-time filling D7-brane. We include non-trivial background fluxes for the D7-brane $U(1)$ field strength on the internal four-cycle wrapped by the brane. First the four-dimensional fermionic spectrum arising from the D7-brane is derived and then the D- and F-terms are obtained by computing appropriate couplings of these fermionic fields. For specific examples we examine the resulting flux-induced scalar potentials and comment on their relevance in string cosmology.

\vfill

\noindent February 2005

\end{titlepage}


\section{Introduction}


Traditionally the contact of string theory with particle physics focused on the low energy limit of the heterotic string. In recent years type~II string compactifications with a stack of space-time filling D-branes have been discussed as possible alternatives \cite{ReviewDB}. In these models the Standard Model or more generally the matter sector 
together with a non-Abelian gauge group 
arises as the massless excitations of the D-branes.
The current paradigm of particle phenomenology prefers an
$\mathcal{N}=1$ supersymmetric matter sector 
spontaneously broken at low energies.
In D-brane models this can be arranged by
compactifying on six-dimensional Calabi-Yau orientifolds
which leave an $\mathcal{N}=1$ supersymmetry unbroken 
\cite{JP,Acharya:2002ag,Brunner:20032004,Grimm:2004}.
This $\mathcal{N}=1$ can be broken spontaneously by additionally
turning on background fluxes in the orientifold bulk \cite{Bachas:1995ik,Polchinski:1995sm,Michelson:1996pn,Gukov:1999ya,Dasgupta:1999ss,Fluxes:19992001,Giddings:2001yu,Haack:2001jz,Becker:2002nn,Giryavets:2003vd,Grimm:2004}. The couplings of the bulk moduli to the D-brane matter fields communicate the breaking of supersymmetry to the standard model sector and soft supersymmetry breaking terms are generated \cite{Grana:2002,Kors:2003wf,Camara:2003ku,Grana:2003ek,Lawrence:2004zk,Lust:2004,Camara:2004jj,Font:2004cx,Lust:2005bd}.

Bulk background fluxes do not only provide for a mechanism to break supersymmetry but
also stabilize neutral moduli fields \cite{Polchinski:1995sm,Fluxes:19992001,Giddings:2001yu,Stabilize:20022004,Cascales:2003pt}. However, generically not all flat directions of the moduli are lifted by background fluxes. The remaining moduli can be fixed, for example, by non-perturbative contributions to the superpotential such as gaugino condensation on a stack of hidden sector D7-branes or by Euclidean D3-brane instantons \cite{KKLT:2003,Denef:2004dm,Gorlich:2004qm,Choi:2004sx}. 

Recently space-time filling D-branes have also been introduced as ingredients in string cosmology \cite{reviewcosmo}. In particular D3-branes and anti-D3-branes in type~IIB Calabi-Yau compactifications can lead to (metastable) deSitter vacua \cite{KKLT:2003}. The simultaneous inclusion of branes and anti-branes breaks supersymmetry explicitly. Alternatively it has been suggested in ref.~\cite{Burgess:2003ic} to replace the anti-D3-branes by D7-branes with internal background fluxes which break supersymmetry spontaneously. These background fluxes are also meant to provide for the positive energy needed for (metastable) deSitter vacua. 

Various aspects of the low energy effective action for 
space-time filling D7-branes which 
wrap a four-cycle of the internal Calabi-Yau orientifold have
been derived in refs.~\cite{Lust:2004cx,Lust:2004,Camara:2004jj,Jockers:2004yj,Lust:2005bd}.
In particular in ref.~\cite{Jockers:2004yj} we employed a
 Kaluza-Klein analysis
of the eight-dimensional D7-brane worldvolume action coupled to 
the ten-dimensional type~IIB bulk supergravity action 
to compute the K\"ahler potential and the gauge kinetic function
of the four-dimensional $\mathcal{N}=1$ supergravity 
in the large volume limit. Furthermore non-trivial two-form fluxes
of the internal gauge field 
on the wrapped four-cycle were turned on. However, in \cite{Jockers:2004yj}
we restricted our attention to a specific class of fluxes 
whose dual two-cycles are also non-trivial two-cycles in the ambient
Calabi-Yau orientifold. Using a supergravity analysis 
we showed that this subclass of fluxes 
contributes to a D-term potential in the effective action in agreement with the general arguments of ref.~\cite{Brunner:1999jq}. However, at the minimum of 
the potential this D-term always vanishes leaving a Minkowskian ground state.

In this paper we extend our previous analysis in two respects.
We consider all possible 
two-form fluxes of the internal gauge field
including fluxes whose dual two-cycles are non-trivial on the four-cycle but
trivial in the ambient Calabi-Yau orientifold.
In addition we 
compute both  D- and F-terms induced by the fluxes.

In \cite{Jockers:2004yj} the D-term was computed somewhat indirectly
by first determining the Killing vector of the gauged isometry
and then infering the D-term using standard supergravity relations.
The same method can be applied to the more general fluxes considered 
in this paper. However, the superpotential cannot be obtained in this
way and one has to employ other methods.
One possibility is the computation of the scalar potential and 
then with the help of the
supergravity relations determine simultaneously the F-term  and the
D-term.  However, for the case at hand this
method is difficult to implement due to the fact that the D- and
F-terms enter in the scalar potential quadratically. A second
possibility is to compute appropriate fermionic couplings which
determine the D- and F-terms directly. We find that this is a powerful
and convenient way to compute couplings in the low energy effective
theory which results in generic expressions for the
D- and F-terms in terms of worldvolume integrals depending
on the non-trivial fluxes.
For D3-branes this method was pioneered in 
ref.~\cite{Grana:2002} while for D7-branes the corresponding computation
has not been spelled out 
in the literature and therefore we spend a considerable amount of time
going through the details of the calculation. 

Our analysis confirms the results of \cite{Jockers:2004yj}  that
fluxes which are non-trivial in the ambient Calabi-Yau manifold amount to a
shift in the D-term which can always be absorbed into a redefinition of a
scalar field. Therefore one always finds a supersymmetric vacuum
in this case.
On the other hand fluxes which are not inherited from the Calabi-Yau generically induce 
a linear superpotential for the D7-brane matter fields. 
This superpotential can also be understood from
the reduction of a holomorphic  Chern-Simons term \cite{Witten:1992fb,Kachru:2000ih,Mayr:2001,Lust:2005bd}.
In special cases these fluxes can also generate
an additional contribution to the D-term as has recently  been
observed in ref.~\cite{Lust:2005bd}.\footnote{We thank the authors of
  ref.~\cite{Lust:2005bd} for communicating their results prior to
  publication.}
In this case the existence
of a stable supersymmetric vacuum depends on topological properties of the 
Calabi-Yau orientifold. Thus constructing a metastable minimum with
spontaneous supersymmetry breaking induced by a D-term 
as suggested in \cite{Burgess:2003ic} is in principle possible but requires a
considerable amount of engineering and also requires 
non-perturbative contributions to the effective potential.
 However, throughout this paper we work at leading order in $\alpha'$. That is to say perturbative string corrections as in refs.~\cite{Becker:2002nn,Balasubramanian:2004uy} are not taken into account.  

The paper is organized as follows. In section~\ref{sec:bosons} we
discuss the bosonic part of the low energy effective action of
Calabi-Yau orientifolds with a space-time filling D7-brane. In order
to set the stage for the forthcoming analysis we first summarize in
section~\ref{sec:review} the results of ref.~\cite{Jockers:2004yj}. In
section~\ref{sec:fluxes} we turn on the D7-brane background fluxes and
show how they affect the definition of the K\"ahler variables and the
K\"ahler potential. We compute the D-term from a supergravity analysis
and discuss the resulting scalar potential. We observe that 
for specific fluxes the $U(1)$ gauge theory becomes anomalous.
This can occur when the D7-brane intersects its orientifold
image and we briefly analyze the anomaly flow at the intersection \cite{Green:1996dd,Cheung:1997az}.

We then turn to the computation of the D- and F-terms via appropriate fermionic couplings in 
section~\ref{sec:fermions}. 
In section~\ref{sec:DBaction} the superspace extension of the Dirac-Born-Infeld and the Chern-Simons action for the space-time filling D7-brane
following \cite{Cederwall:1996,Bergshoeff:1996tu} is reviewed. 
In particular we discuss how the fermionic brane degrees of freedom are encoded into the superspace D7-brane action.
This in turn allows us  
to determine in section~\ref{sec:fspec}
the massless four-dimensional fermionic spectrum which completes the vector and chiral multiplets whose bosonic components 
were introduced in \ref{sec:review}.
In  section~\ref{sec:fsugra} we briefly review the fermionic part of the $\mathcal{N}=1$ supergravity action and identify the 
gravitino couplings 
which determine the D- and F-terms. However, as a first step it is
necessary to compute their kinetic terms, which enables us to 
adjust the normalization of the fermionic D-brane excitations 
with the supergravity normalization.
Section~\ref{sec:DFterms} contains our main result in that we compute 
certain fermionic couplings and determine a generic expression for the
D-term and the superpotential in terms of D7-brane worldvolume integrals
depending on the fluxes. Finally in section~\ref{sec:HCS} we argue that the derived superpotential can also be obtained by dimensional reduction from the holomorphic Chern-Simons action.

In section~\ref{sec:scalarpot} we examine two instructive examples in
more detail
and determine the structure of their flux-induced scalar
potentials. We comment on the relevance of these models for
cosmological applications along the lines of
refs.~\cite{KKLT:2003}. In section~\ref{sec:onet} the first model with
one geometric bulk modulus parametrizing the volume of the internal
Calabi-Yau orientifold is considered. 
The second example in section~\ref{sec:threet} is closely related to toroidal compactifications \cite{Lust:2004cx,Lust:2004,Lust:2005bd}. 

Section~\ref{sec:conc} contains our conclusions and notation and some of the technical details are assembled in two appendices. In appendix~\ref{sec:action} we give the bosonic low energy effective action for the Calabi-Yau orientifold compactification with a D7-brane as derived in \cite{Jockers:2004yj} and our conventions for the fermionic fields and their Dirac gamma matrices are summarized in appendix~\ref{sec:conv}. 


\section{Bosonic action for D7-branes in $\mathcal{N}=1$ Calabi-Yau orientifolds} 
\label{sec:bosons}


In order to set the stage we briefly summarize in section~\ref{sec:review} the results of ref.~\cite{Jockers:2004yj} and recall the low energy effective description of Calabi-Yau orientifold compactifications with space-time filling D7-branes. Specifically we review the bosonic spectrum, the K\"ahler variables and the K\"ahler potential for the resulting supergravity theory. In  section~\ref{sec:fluxes} we introduce D7-brane background fluxes and derive the change in the complex structure of the target space K\"ahler manifold.


\subsection{Calabi-Yau orientifold compactifications with D7-branes} \label{sec:review}


The starting point is type~IIB string theory compactified on a Calabi-Yau orientifold. To be specific the ten-dimensional space-time background $M^{9,1}$ is taken to be the product 
\begin{equation}\label{eq:Pansatz}
   M^{9,1}= \mathbb{R}^{3,1}\times Y/\mathcal{O}\ ,
\end{equation}
 where the internal Calabi-Yau orientifold $Y/\mathcal{O}$ is a compact Calabi-Yau manifold~$Y$ moded out by  a discrete involutive symmetry $\mathcal{O}$. The corresponding metric takes the form
\begin{equation} \label{eq:met}
   \dd s_{10}^2 = \hat\eta_{\mu\nu}\:\dd x^\mu\dd x^\nu+
                  2\:\hat g_{i\bar\jmath}(y)\:\dd y^i \dd \bar y^{\bar\jmath} \ ,
\end{equation}
where $\hat\eta_{\mu\nu}$ is the metric of the four-dimensional Minkowski space and $\hat g_{i\bar\jmath}(y)$ is the metric of the internal Calabi-Yau manifold~$Y$.\footnote{The hat $\hat{}$ denotes quantities in the string frame.}
In the presence of $\mathcal{N}=1$ supersymmetric  space-time filling D7-branes 
consistency requires to also introduce space-time filling O3/O7-planes
preserving the same supercharge. This in turn determines 
the orientifold projection $\mathcal{O}$ to be \cite{Vafa:1995gm,Dabholkar:1996pc,Acharya:2002ag,Brunner:20032004}
\begin{equation} \label{eq:proj}
   \mathcal{O}=(-1)^{F_L}\Omega_p \sigma^* \ ,\qquad
\mathcal{O}^2 = \id\ ,
\end{equation}
where $F_L$ is the fermion number for the left-movers and $\Omega_p$ is the world-sheet parity operator. $\sigma$ is an isometric, holomorphic involution on the Calabi-Yau manifold $Y$ with  O3/O7 planes as its fixed point locus.
Furthermore, it acts  via pullback on the type IIB fields and satisfies 
the additional property \cite{Acharya:2002ag,Brunner:20032004}
\begin{equation} \label{eq:invOmega}
   \sigma^*\Omega=-\Omega \ ,\qquad (\sigma^*)^2 = \id \ ,
\end{equation}
where $\Omega$ denotes the unique $(3,0)$-form of the Calabi-Yau manifold~$Y$.

Before we continue let us discuss the validity of our ansatz~\eqref{eq:Pansatz}. For Calabi-Yau compactifications with localized sources such as orientifold planes and/or D-branes one really has to make a warped ansatz for the metric and include a dilaton gradient in order to capture the back-reaction to geometry \cite{Giddings:2001yu,DeWolfeDeAlwisBuchel}. However, if the D-brane tension is cancled locally by the negative tension of orientifold planes, that is to say if D-branes are on top of the orientifold planes, there is no backreaction to geometry. Therefore in the following we perform the analysis in the regime, where the internal space is large such that the warp factor becomes constant and where the separation of the D-brane from the orientifold planes is small enough such that the dilaton gradient is small. In this regime the back-reaction should be viewed as a perturbation to the product ansatz, which generates in the effective action scalar potential terms. It would be interesting to study the potentials arising from this backreaction, which, however, is beyond the scope of this paper.

\begin{table}
\begin{center}
\begin{tabular}{|c|c|c||c|c|c|}
   \hline
      \bf space  &  \bf basis  &  \bf dimension  &
      \bf space  &  \bf basis  &  \bf dimension  
      \rule[-1.5ex]{0pt}{4.5ex} \\
   \hline
   \hline
      $H^{(1,1)}_{\bar\partial,+}(Y)$  &  $\omega_\alpha$
      &  $\alpha=1,\ldots,h^{1,1}_+$ 
      &  $H^{(1,1)}_{\bar\partial,-}(Y)$  &  $\omega_a$
      &  $a=1,\ldots,h^{1,1}_-$
      \rule[-1.5ex]{0pt}{4.5ex} \\
   \hline
      $H^{(2,2)}_{\bar\partial,+}(Y)$  &  $\tilde\omega^\alpha$
      &  $\alpha=1,\ldots,h^{2,2}_+$ 
      &  $H^{(2,2)}_{\bar\partial,-}(Y)$  &  $\tilde\omega^a$
      &  $a=1,\ldots,h^{1,1}_-$
      \rule[-1.5ex]{0pt}{4.5ex} \\
   \hline
      $H^{3}_+(Y)$  &  $\alpha_{\hat\alpha},\beta^{\hat\alpha}$ 
      &  $\hat\alpha=1,\ldots,h^{2,1}_+$
      &  $H^{3}_-(Y)$  &  $\alpha_{\hat a},\beta^{\hat a}$ 
      &  $\hat a=0,\ldots,h^{2,1}_-$
      \rule[-1.5ex]{0pt}{4.5ex} \\
    \hline
      $H^{(2,1)}_{\bar\partial,+}(Y)$  &  $\chi_{\tilde\alpha}$  
      &  $\tilde\alpha=1,\ldots,h^{2,1}_+$
      &  $H^{(2,1)}_{\bar\partial,-}(Y)$  &  $\chi_{\tilde a}$
      &  $\tilde a=1,\ldots,h^{2,1}_-$
      \rule[-1.5ex]{0pt}{4.5ex} \\
    \hline  
      $H^{(1,2)}_{\bar\partial,+}(Y)$  &  $\bar\chi_{\tilde\alpha}$  
      &  $\tilde\alpha=1,\ldots,h^{2,1}_+$
      &  $H^{(1,2)}_{\bar\partial,-}(Y)$  &  $\bar\chi_{\tilde a}$
      &  $\tilde a=1,\ldots,h^{2,1}_-$
      \rule[-1.5ex]{0pt}{4.5ex} \\
    \hline  
\end{tabular} 
\caption{Cohomology basis} \label{tab:coh} 
\end{center}
\end{table}
The massless fields of the orientifold Calabi-Yau compactification are obtained by performing a Kaluza-Klein reduction of the ten-dimensional supergravity spectrum keeping only zero modes which are invariant under the orientifold projection \eqref{eq:proj}.
Such zero modes are in one-to-one correspondence with
harmonic forms of the Calabi-Yau manifold~$Y$
and thus determined by the cohomology groups $H^{(p,q)}_{\bar\partial}(Y)$.
Due to the holomorphicity and the involutive property of $\sigma$ the groups
$H^{(p,q)}_{\bar\partial}(Y)$ split into
even and odd eigenspaces
\begin{equation}
   H^{(p,q)}_{\bar\partial}(Y) =
   H^{(p,q)}_{\bar\partial,+}(Y)\oplus H^{(p,q)}_{\bar\partial,-}(Y)\ ,
\end{equation}
which we list together with their respective basis elements 
in Table~\ref{tab:coh}.

Expanding the massless ten-dimensional fields of type IIB string theory
in terms of harmonic forms keeping only the $\mathcal{O}$ invariant 
four-dimensional modes results in the massless $\mathcal{N}=1$ bulk spectrum 
\cite{Acharya:2002ag,Brunner:20032004,Grimm:2004}.
For the ten-dimensional NS-NS fields that is to say for the dilaton $\phi$, the metric $g$ and the two-form $B$ one finds the expansion \cite{Grimm:2004}
\begin{align} \label{eq:NS}
   J&=v^\alpha(x)\:\omega_\alpha \ , & B&=b^a(x)\:\omega_a \ , & \phi&=\phi(x) \ ,
\end{align}
where $J$ is the K\"ahler-form of the Calabi-Yau manifold $Y$
and $v^\alpha(x), b^a(x)$ and $\phi(x)$ are four-dimensional scalar 
fields.\footnote{For ease of notation we denote in the following both the Calabi-Yau orientifold and the Calabi-Yau manifold by $Y$.} 
In addition deformations of the complex structure on $Y$ lead to 
complex scalars $z^{\tilde a}$ which are in one-to-one correspondence with the elements of $H_{\bar\partial,-}^{(2,1)}(Y)$ \cite{Brunner:20032004,Candelas:1990pi}. 

Similarly, the ten-dimensional RR form fields~$C^{(0)}$, $C^{(2)}$ and $C^{(4)}$ are expanded into appropriate harmonic forms as
\begin{equation} \label{eq:C}
\begin{split}
   C^{(4)}&=D_{(2)}^\alpha(x)\wedge\omega_\alpha
            +V^{\hat\alpha}(x)\wedge \alpha_{\hat\alpha}
            +U_{\hat\alpha}(x)\wedge \beta^{\hat\alpha}
            +\rho_\alpha(x)\:\tilde\omega^\alpha \ , \\
   C^{(2)}&=c^a(x)\:\omega_a \ ,\qquad\qquad\qquad C^{(0)}=l(x) \ ,
\end{split}
\end{equation}
where $D_{(2)}^\alpha$ are two-forms, 
$V^{\hat\alpha}, U_{\hat\alpha}$ are one-forms and $\rho_\alpha,c^a,l$
are scalars in $d=4$. The 
 physical spectrum is obtained by imposing the self-duality condition on the five-form field strength, which removes half of the degrees of freedom of $C^{(4)}$. This can be used to eliminate the two-form fields $D_{(2)}^\alpha$ in favor of the scalars $\rho_\alpha$ and the vectors $V^{\hat\alpha}$ in favor of the vectors $U_{\hat\alpha}$. The resulting physical spectrum without redundant degrees of freedom is summarized in Table~\ref{tab:sp}. 
\begin{table}
\begin{center}
\begin{tabular}{|c|c|c||c|c|c|}
   \hline
      \bf multiplet  &  \bf multiplicity &  \bf bos. fields &
      \bf multiplet  &  \bf multiplicity &  \bf bos. fields \rule[-1.5ex]{0pt}{4.5ex} \\
   \hline
   \hline
      gravity  &  $1$  &  $g_{\mu\nu}$  &
      chiral &  $h_-^{1,1}$  &  $(b^a,c^a)$ \rule[-1.5ex]{0pt}{4.5ex} \\
   \hline
      vector  &  $h_+^{2,1}$  &  $V^{\hat\alpha}_\mu$  & 
      chiral  &  $h_+^{1,1}$  &  $(\rho_\alpha,v^\alpha)$ \rule[-1.5ex]{0pt}{4.5ex} \\
   \hline
      chiral &  $1$  &  $(l,\phi)$  &
      chiral &  $h_-^{2,1}$  &  $z^{\tilde a}$ \rule[-1.5ex]{0pt}{4.5ex} \\
   \hline
\end{tabular} 
\caption{$\mathcal{N}=1$ bulk multiplets} \label{tab:sp} 
\end{center}
\end{table}

The next task is to add to the orientifold bulk theory a single space-time filling D7-brane with gauge group~$U(1)$.\footnote{Note that a D7-brane cannot be included into the bulk theory arbitrarily. Instead in order to obtain a consistent theory the tadpole cancellation conditions for branes and orientifold planes must be satisfied \cite{StefanskiScrucca,Blumenhagen:2002wn}. We come back to this issue in section~\ref{sec:fluxes}.} The internal part of the D7-brane is wrapped on a four cycle~$S^\Lambda$ which includes both the D7-brane cycle and its image with respect to the orientifold involution~$\sigma$. Therefore the cycle~$S^\Lambda$ describes two disconnected components and the two dimensional cohomology space $H^0(S^\Lambda)$ decomposes into the two one dimensional spaces $H^0_+(S^\Lambda)$ with basis element
$1$, and $H^0_-(S^\Lambda)$ with the basis element
\begin{equation} \label{eq:P}
   P_-\in H^0_-(S^\Lambda) \ .
\end{equation}
The zero form~$P_-$ is $+1$ on the D7-brane cycle and $-1$ on the image-D7-brane cycle.  
For later convenience we also introduce the four cycle~$S^P$, which is the union of the D7-brane cycle and its orientation reversed image under the involution~$\sigma$, i.e. these two cycles obey
\begin{align}\label{eq:SP}
  \sigma(S^\Lambda)=S^\Lambda \ , && \sigma(S^P)=-S^P \ .
\end{align}
Their Poincar\'e dual two-forms are denoted by $\omega_\Lambda\in H^{(1,1)}_{\bar\partial,+}(Y)$ and $\omega_P\in H^{(1,1)}_{\bar\partial,-}(Y)$ respectively.

The massless bosonic spectrum resulting from the space-time filling D7-brane wrapped on $S^\Lambda$ consists of a four-dimensional 
$U(1)$ gauge field $A_\mu(x)$ and Wilson line moduli fields $a_I(x)$ both arising
from the eight-dimensional world-volume gauge field. Furthermore,
fluctuations of the internal cycle $S^\Lambda$ lead to 
`matter fields' $\db^A(x)$ which arise from a normal coordinate expansion of the D7-brane.
In the limit of small D7-brane fluctuations~$\db^A$ and small complex structure deformations~$z^{\tilde a}$ these fields can be treated independently. As a consequence the `matter fields'~$\db^A$ appear as an expansion into two-forms of $S^\Lambda$ of type $(2,0)$ \cite{Jockers:2004yj}. The massless spectrum is summarized in Table~\ref{tab:spec} together with their associated basis of harmonic forms.
\begin{table}
\begin{center}
\begin{tabular}{|c|c|c|c|}
   \hline
      \bf multiplet  &  \bf bosonic fields  
         &  \bf geometric space &  \bf basis \rule[-1.5ex]{0pt}{4.5ex} \\
   \hline
   \hline
  vector &  $A_\mu$
         &  $H_-^0(S^\Lambda)$   &  $\{P_-\}$ 
         \rule[-1.5ex]{0pt}{4.5ex} \\
   \hline
  chiral Wilson lines &  $a_I$, $I=1,\ldots, \dim H_{\bar\partial,-}^{(0,1)}(S^\Lambda)$
         &  $H_{\bar\partial,-}^{(0,1)}(S^\Lambda)$   &  $\{A^I\}$ 
         \rule[-1.5ex]{0pt}{4.5ex} \\
   \hline
  chiral matter  &  $\db^A$, $A=1,\ldots, \dim H_{\bar\partial,-}^{(2,0)}(S^\Lambda)$ 
          &  $H_{\bar\partial,-}^{(2,0)}(S^\Lambda)$  &  $\{\tilde s_A\}$ 
         \rule[-1.5ex]{0pt}{4.5ex} \\
   \hline
\end{tabular} 
\caption{Massless D7-brane spectrum} \label{tab:spec} 
\end{center}
\end{table}

In the large volume limit the bosonic four-dimensional low energy effective supergravity action is derived 
by a Kaluza-Klein reduction on the orientifold background from the ten-dimensional type~IIB supergravity action 
plus the Dirac-Born-Infeld and Chern-Simons action integrated over the D7-brane worldvolume $\mathcal{W}=\mathbb{R}^{(3,1)}\times S^\Lambda$ 
\cite{Grimm:2004,Jockers:2004yj}. 
 The result of this compactification procedure yields a $\mathcal{N}=1$ supergravity theory in four dimensions in terms of the vector fields and chiral fields specified in Table~\ref{tab:sp} and \ref{tab:spec}.
The bosonic low energy effective action obtained 
from this reduction is recorded in 
\eqref{eq:action}.
Here we only give the $\mathcal{N}=1$ supergravity action 
in its standard form, that is we specify 
 the K\"ahler potential~$K$, the holomorphic superpotential~$W$ and the holomorphic gauge kinetic coupling functions~$f_{\Gamma\Delta}$.
In terms of these quantities the action is  given by  \cite{Cremmer:1982en,Wess:1992}
\begin{equation} \label{eq:4Dbos}
\begin{split}
   \mathcal{S}_\text{Bosons}=&-\frac{1}{2\kappa_4^2}\int\dd^4x\:\sqrt{-\eta}\left( R 
      + 2\:K_{M\bar N}\nabla_\mu M^M\nabla^\mu \bar M^{\bar N} + 
        2\:V_\text{D} +2\:V_\text{F} \right) \\
     &-\frac{1}{4\kappa_4^2}\int\dd^4x\:\sqrt{-\eta}\:(\Real f)_{\Gamma\Delta}
        F_{\mu\nu}^\Gamma F^{\mu\nu\:\Delta} 
      + \frac{1}{2\kappa_4^2}\int(\Imag f)_{\Gamma\Delta} F^\Gamma\wedge F^\Delta \ ,
\end{split}
\end{equation}
where we denoted all scalar fields of the chiral multiplets collectively by 
$M^M$ and all gauge fields and their field strength by 
$V^\Gamma, F^\Gamma$ respectively.
$K_{M\bar N}=\partial_M\partial_{\bar N}K$ is the K\"ahler metric and  the scalar potential is the sum of 
the two terms 
\begin{align} \label{eq:spot}
   V_\text{F}=e^K\left(K^{M\bar N}\mathcal{D}_M W\mathcal{D}_{\bar N}\bar W -3|W|^2\right) \ , &&
   V_\text{D}=\frac{1}{2}\left(\Real{f}\right)^{\Gamma\Delta} \text{D}_\Gamma\text{D}_\Delta \ ,
\end{align}
where $\mathcal{D}_M W=\partial_M W + \left(\partial_MK\right) W$ 
and $\left(\Real f\right)^{\Gamma\Delta}$ is the inverse matrix
of the real part of the coupling matrix~$f_{\Gamma\Delta}$. 

As demonstrated in ref.~\cite{Grimm:2004,Jockers:2004yj} in order to cast the low energy effective action obtained from the Kaluza-Klein reduction 
and given in \eqref{eq:action} into the form \eqref{eq:4Dbos}
the K\"ahler variables must be identified, i.e. the correct complex structure of the space spanned by the chiral fields must be determined. The result of this analysis shows that $S$, $G^a$, $T_\alpha$, $z^{\tilde a}$, $\db^A$ and $a_I$ are the appropriate K\"ahler variables, where $S$, $G^a$, $T_\alpha$ are defined as
\begin{align}
   \dil&=\tau-\kappa_4^2\mu_7\mathcal{L}_{A\bar B}\db^A\bar\db^{\bar B}  \ , \qquad\qquad
   G^a=c^a-\tau b^a \ , \label{eq:G} \\ 
   T_\alpha&=\frac{3i}{2}\left(\rho_\alpha
      -\tfrac{1}{2}\mathcal{K}_{\alpha bc}c^b b^c\right) +\frac{3}{4}\mathcal{K}_\alpha
      +\frac{3i}{4(\tau-\bar\tau)} \mathcal{K}_{\alpha bc}G^b(G^c-\bar G^c)
      +3i\kappa_4^2\mu_7\ell^2 \mathcal{C}^{I\bar J}_\alpha a_I \bar a_{\bar J} \ ,\nonumber
\end{align}
with $\ell=2\pi\alpha'$ and $\tau=l+ie^{-\phi}$. The field independent coefficients $\mathcal{L}_{A\bar B}$, $\mathcal{C}_{\alpha}^{I\bar J}$ are defined in \eqref{eq:LC}, whereas the triple intersection numbers $\mathcal{K}_{\alpha bc}$ and $\mathcal{K}_\alpha$ are defined in \eqref{eq:triple} and \eqref{eq:K}.

In terms of these K\"ahler coordinates the K\"ahler potential for the low energy effective supergravity action is found to be \cite{Jockers:2004yj}
\begin{multline} \label{eq:K1}
   K(\dil, G, T, z, \db, a)=K_\text{CS}(z) 
   -\log\left[-i\left(\dil-\bar\dil\right)-2i\kappa_4^2\mu_7\mathcal{L}_{A\bar B}
   \db^A\bar\db^{\bar B}\right] \\
     -2 \log\left[\tfrac{1}{6}\mathcal{K}(\dil,G,T,\db,a)\right] \ ,
\end{multline}
where $K_\text{CS}(z)$ is the K\"ahler potential of the complex structure moduli~$z^{\tilde a}$ defined in eq.~\eqref{eq:CSt}. 
$\mathcal{K} =\mathcal{K}_{\alpha\beta\gamma} v^\alpha v^\beta v^\gamma$
is known explicitly 
in terms of the Kaluza-Klein variables
$v^\alpha$ arising in the expansion of $J$ (c.f.~\eqref{eq:NS}) 
and the triple intersections
$\mathcal{K}_{\alpha\beta\gamma}$ defined in eq.~\eqref{eq:triple}.
The $v^\alpha$ themselves are no   K\"ahler coordinates but 
determined in terms of the K\"ahler coordinates $S$, $G^a$, $T_\alpha$, $\db^A$ and $a_I$ by solving \eqref{eq:G} for $v^\alpha(S,G,T,\db,a_I)$
\cite{Haack:1999zv,Becker:2002nn,Grana:2003ek,Grimm:2004}. This solution, however, cannot be given explicitly in general.

The gauge kinetic coupling function of the D7-brane gauge degrees of freedom is extracted from the effective action \eqref{eq:action}\footnote{Actually there is a slight mismatch involving the Wilson line moduli fields~$a_I$ \cite{Hsu:2003cy}, which is cured at the open string one loop level \cite{Berg:2004ek}.} 
\begin{equation} \label{eq:f}
   f^\text{D7}=\frac{2\kappa_4^2\mu_7\ell^2}{3}\:T_\Lambda \ .
\end{equation}

The computation of the scalar potential from a Kaluza-Klein reduction
is more delicate. The reason is that in the reduction we used 
explicitly the BPS-condition for the D7-brane or in other words
we wrapped it on a supersymmetric cycle. This amounts to choosing
a flat $\mathcal{N}=1$ supersymmetric background and as a consequence
no potential can appear. It was argued in ref.~\cite{Jockers:2004yj} that 
the deviation from the BPS condition can be viewed as inducing 
a D-term. However, the precise computation of this D-term in the bosonic 
action is difficult.
We will return to this issue in more detail 
in the next section and here only observe that 
the D-term can also be computed from a supergravity analysis.
As reviewed in 
appendix~\ref{sec:action} the action \eqref{eq:action} has 
a local Peccei-Quinn symmetry \eqref{eq:shift} 
under which one of the scalars $G^a$  defined in 
\eqref{eq:G} is charged. This scalar denoted by $G^P$ arises
from the expansion along the (1,1)-form $\omega_P$ which is dual
to the four-cycle $S^P$ defined in \eqref{eq:SP}.
Using \eqref{eq:G} and \eqref{eq:cd1} one determines the gauge covariant derivative to be
\begin{equation} \label{eq:Gcov}
   \nabla_\mu G^P=\partial_\mu G^P - 4\kappa_4^2\mu_7\ell A_\mu \ .
\end{equation}
The D-terms associated to the charged chiral fields are in general computed from the equation \cite{Wess:1992}
\begin{equation} \label{eq:Dterms}
   \partial_N\partial_{\bar M}K\:\bar X^{\bar M}_\Gamma=i\partial_N\text{D}_\Gamma \ ,
\end{equation}
where $X_\Gamma$ is the holomorphic Killing vector field of the corresponding gauged isometry of the target space K\"ahler manifold. For the shift symmetry \eqref{eq:Gcov} the Killing vector of the gauged isometry is easily determined to be
\begin{equation} \label{eq:Killing1}
   X=4\kappa_4^2\mu_7\ell \partial_{G^P} \ .
\end{equation}
Then using \eqref{eq:Dterms} we readily compute the D-term associated to this non-linearly realized $U(1)$ gauge symmetry to be
\begin{equation} \label{eq:D}
   \text{D}=\frac{12\kappa_4^2\mu_7\ell}{\mathcal{K}}\
  {\mathcal{K}_{Pa}b^a}\ ,
\end{equation} 
where $\mathcal{K}_{Pa}$ is defined in \eqref{eq:K}.
Using \eqref{eq:NS} we can also give an integral representation
for the D-term which reads
\begin{equation}
   \label{eq:Di}
   \text{D}=\frac{12\kappa_4^2\mu_7\ell}{\mathcal{K}} \int_{S_P} J\wedge B \ .
\end{equation}
Moreover with eq.~\eqref{eq:spot} the corresponding D-term scalar potential $V_\text{D}$ becomes
\begin{equation}
   V_\text{D}=\frac{108\kappa_4^2\mu_7}{\mathcal{K}^2\Real{T_\Lambda}}
     \left(\mathcal{K}_{Pa}b^a\right)^2 \ .
\end{equation}
$V_\text{D}$ is minimized for $b^a=0$ where the D-term and $V_\text{D}$ itself
vanish.
This concludes the summary of the ingredients needed in the following chapters. The details of this section are elaborated in ref.~\cite{Jockers:2004yj}.


\subsection{D7-brane background fluxes} \label{sec:fluxes}


In this section we turn on background fluxes for the field strength of the $U(1)$ gauge theory localized on the D7-brane worldvolume. However, in order to preserve Poincar\'e invariance of the four-dimensional effective theory we consider only background fluxes on the internal D7-brane cycle~$S^\Lambda$. 
These fluxes are topologically non-trivial two-form configurations for the 
internal $U(1)$ field strength which nevertheless satisfy
the Bianchi identity and the equation of motion. Therefore
the background flux $f$ is constrained to be a harmonic form on $S^\Lambda$.
In ref.~\cite{Jockers:2004yj} it was shown that 
 the gauge boson is odd with respect to the orientifold involution~$\sigma$ 
and as a consequence the background flux~$f$ has to be
 an element of $H_-^2(S^\Lambda)$. 

Let us now pause to briefly discuss the tadpole cancellation conditions. Consistency requires the cancellation of all RR~tadpoles. In the case of a single D7-brane wrapped on the internal cycle~$S^\Lambda$ and with D7-brane flux~$f$, the RR~tadpole cancellation conditions read \cite{Blumenhagen:2002wn}
\begin{equation} \label{eq:tadpoles}
\begin{split}
   0&= \mu_7\int_{\mathbb{R}^{3,1}\times S^\Lambda} C^{(8)}
      +\sum_j \nu_7^j \int_{\mathbb{R}^{3,1}\times O_j^{(7)}} C^{(8)} \ , \\
   0&= \mu_7\ell^2\int_{\mathbb{R}^{3,1}\times S^\Lambda} C^{(4)}\wedge f \wedge f
      +\sum_l \nu_3^l \int_{\mathbb{R}^{3,1}} C^{(4)} \ .
\end{split}
\end{equation}
Here $\nu_7^j$ and $\nu_3^l$ are the RR~charges of the O7- and O3-planes respectively whereas $O_j^{(7)}$ are the internal four-cycles wrapped by the O7-planes. Since we do not analyze a specific orientifold compactification but instead work with a generic ansatz, we cannot check the conditions \eqref{eq:tadpoles} explicitly. Therefore we assume that in the following the Calabi-Yau manifold~$Y$, the involution~$\sigma$, the D7-brane cycle~$S^\Lambda$ and the D7-brane flux~$f$ is chosen in such a way that the conditions~\eqref{eq:tadpoles} are fulfilled.

In addition to RR~tadpoles there can also appear NS-NS~tadpoles. The divergencies of NS-NS tadpoles give rise to potentials for the NS-NS fields \cite{Dudas:2000ff,Blumenhagen:2001te} and can be absorbed in the background fields via the Fischler-Susskind mechanism \cite{Fischler:1986}. In the effective theory the presence of NS-NS tadpoles generically indicate an unstable background. However, in the supersymmetric case the RR~tadpole conditions~\eqref{eq:tadpoles} imply that also the NS-NS~tadpoles vanish via supersymmetry. If the NS-NS~tadpoles do not vanish a D-term in the effective action is induced \cite{Blumenhagen:2001te}, which breaks supersymmetry spontaneously and generically indicates that the vacuum expectation values of the effective four-dimensional fields do not correspond to a minimum in the scalar potential.

The D7-brane cycle $S^\Lambda$ is embedded into the ambient Calabi-Yau manifold~$Y$ via the embedding map $\iota:S^\Lambda\hookrightarrow Y$, which induces the pullback map $\iota^*$ on forms
\begin{equation}
   \iota^*:H_-^2(Y)\rightarrow H_-^2(S^\Lambda) \ .
\end{equation}
Therefore one can distinguish between two different kinds of fluxes
which we denote by $\leftup{Y}{f}$ and $\tilde f$.
$\leftup{Y}{f}$ are harmonic two-forms on $S^\Lambda$ which are 
inherited from the ambient Calabi-Yau space~$Y$.
 $\tilde f$ on the other hand correspond to harmonic forms on $S^\Lambda$, which cannot be obtained by pullback from the ambient space~$Y$.
Put differently, $\leftup{Y}{f}$ are harmonic two-forms in the image of $\iota^*$
while $\tilde f$ are harmonic two-forms in the cokernel of $\iota^*$.
This amounts to the fact that the cohomology group $H_-^2(S^\Lambda)$ 
can be decomposed 
\begin{equation} \label{eq:FluxCoh}
   H_-^2(S^\Lambda)\cong \leftup{Y}{H_-^2(S^\Lambda)}\oplus \tilde H_-^2(S^\Lambda) \ ,
\end{equation}
where $\leftup{Y}{H_-^2(S^\Lambda)}=\iota^*\left(H_-^2(Y)\right)$ and $\tilde H_-^2(S^\Lambda)=\coker\left(H_-^2(Y)\xrightarrow{\iota^*} H_-^2(S^\Lambda)\right)$. 
Then the flux $f\in H_-^2(S^\Lambda)$ splits accordingly 
\begin{equation}\label{eq:fsplit}
   f=\leftup{Y}{f}+\tilde f \ ,
\end{equation} 
with $\leftup{Y}{f} \in \leftup{Y}{H_-^2(S^\Lambda)}$ and $\tilde f \in \tilde H_-^2(S^\Lambda)$. This splitting is not unique but we choose it in such a way that the integrals 
\begin{equation} \label{eq:vanishint}
   \int_{S^\Lambda}\iota^*\omega_a \wedge \tilde f=0 
\end{equation}
vanish for all two-forms $\omega_a$ in $H^2_-(Y)$.\footnote{This can alway be achieved 
by first choosing a basis of two-forms~$\omega_a$ for $\leftup{Y}H_-(S^\Lambda)$ and then by choosing a basis of two-cycles~$\tilde S^{\tilde a}$ for $\ker\: (H_{2,-}(S^\Lambda)\xrightarrow{\iota_*}H_{2,-}(Y))$. Then the Poincar\'e dual basis~$\omega_{\tilde a}$ of $\tilde S^{\tilde a}$ spans the cokernel $\tilde H^2_-(S^\Lambda)$. If now the splitting is chosen in such a way that $\leftup{Y}{f}$ can be expanded into $\omega_a$ and $\tilde f$ into $\omega_{\tilde a}$ the relation~\eqref{eq:vanishint} is fulfilled.}

Let us first discuss the fluxes $\leftup{Y}{f}$.
In the Calabi-Yau threefold~$Y$ the only harmonic two-forms are $(1,1)$-forms  and therefore the fluxes $\leftup{Y}{f}$ can be expanded into the $(1,1)$-forms of $Y$ pulled back to $S^\Lambda$, namely
\begin{equation} \label{eq:f1}
    \leftup{Y}{f}=\leftup{Y}{f}^a\:\iota^*\omega_a \ ,
\end{equation}
where $\omega_a$ is a basis of $H^{(1,1)}_{\bar\partial,-}(Y)$.
These fluxes have already been treated in ref.~\cite{Jockers:2004yj}
and always appear 
in the combination $b^a-\ell\:\leftup{Y}{f}^a$ in the effective action 
\eqref{eq:faction}. This is due to the fact that the D7-brane
effective action \eqref{eq:DBI} and \eqref{eq:CS} depends on
$\mathcal{F}\equiv B-\ell f$. 
Rewriting \eqref{eq:faction} in the standard $\mathcal{N}=1$ form 
\eqref{eq:4Dbos} amounts to a modification of the chiral coordinates~$G^a$
in that the definition \eqref{eq:G} is replaced by 
\begin{equation} \label{eq:Gf}
   G^a=c^a-\tau\dbbf^a \ , \qquad \dbbf^a=b^a-\ell\:\leftup{Y}{f}^a \ .
\end{equation}
In terms of this newly defined  $G^a$ 
the K\"ahler potential \eqref{eq:K1} and the covariant derivative 
\eqref{eq:Gcov} are unchanged.  As in the previous section
the D-term can be computed from a supergravity analysis
and one finds that \eqref{eq:D} is replaced by
\begin{equation} \label{eq:D2}
   \text{D}\ =\ \frac{12\kappa_4^2\mu_7\ell}{\mathcal{K}}\ 
{\mathcal{K}_{Pa}\dbbf^a}\ =\ 
\frac{12\kappa_4^2\mu_7\ell}{\mathcal{K}}\int_{S_P} J\wedge \dbbf\ ,
\end{equation} 
where $\dbbf = \dbbf^a\omega_a$.
Thus the presence of the fluxes $\leftup{Y}{f}^a$ shifts the vacuum
in that 
$V_\text{D}$ is now minimized for $\dbbf = 0$ which determines
$b^a=\ell\:\leftup{Y}{f}^a$ 
corresponding to a
vanishing D-term $\text{D}=0$.

Let us now turn to the second type of fluxes $\tilde f$ introduced in
\eqref{eq:fsplit}. In a Kaluza-Klein reduction of the D7-brane action
they arise from the same combination $\mathcal{F} \equiv B-\ell f$.  However, due to \eqref{eq:vanishint} some of the integrals encountered simplify.  This implies that no cross terms between $\tilde f$ and any bulk scalars can survive and in the effective action terms  proportional to
\begin{equation}
   \fcharge=\ell^2\int_{S^\Lambda} \tilde f\wedge\tilde f \ 
\end{equation}
appear. 

The explicit form of the Kaluza-Klein reduced effective action
including both types of fluxes $\leftup{Y}{f}^a$ and $\tilde f$
is given in \eqref{eq:faction}. 
In the $\mathcal{N}=1$ supergravity language the additional terms proportional to $\fcharge$ amount to an adjustment of the K\"ahler coordinate $T_\Lambda$ and the definition \eqref{eq:G} is replaced by 
\begin{multline} \label{eq:Tf}
   T_\alpha=\frac{3i}{2}\left(\rho_\alpha
      -\tfrac{1}{2}\mathcal{K}_{\alpha bc}c^b\dbbf^c\right) +\frac{3}{4}\mathcal{K}_\alpha 
      +\frac{3i}{4(\tau-\bar\tau)} \mathcal{K}_{\alpha bc}G^b(G^c-\bar G^c) \\ 
      +3i\kappa_4^2\mu_7\ell^2 \mathcal{C}^{I\bar J}_\alpha a_I \bar a_{\bar J} 
      +\tfrac{3i}{4}\:\delta_\alpha^\Lambda\:\tau\fcharge \  ,
\end{multline}
where $\dbbf^c$ is defined in \eqref{eq:Gf}. 
The K\"ahler potential \eqref{eq:K1} is unchanged but 
the additional terms proportional to $\fcharge$ enter nevertheless once $K$ is expressed in terms of its chiral coordinates.

The covariant derivatives of the scalars $\rho_\alpha$ also change.
This can be seen from the fact that  the Kaluza-Klein reduction of the 
D7-brane Chern-Simons action \eqref{eq:CS}
in the presence of the fluxes~$\tilde f$ 
induces additional Green-Schwarz terms of the form
\begin{equation}\label{eq:GSterms}
   \mu_7\int_\mathcal{W} C^{(4)}\wedge\ell\tilde f\wedge\ell P_-F
   = \mu_7\ell \,Q_\alpha \int_{\mathbb{R}^{3,1}} D^\alpha\wedge F 
   = -\mu_7\ell \,Q_\alpha \int_{\mathbb{R}^{3,1}} \dd D^\alpha\wedge A \ .
\end{equation}
Here we used the expansion \eqref{eq:C} and $Q_\alpha$ is defined as
\begin{equation} \label{eq:notvanish}
   Q_\alpha=\ell \int_{S^\Lambda} \iota^*\omega_\alpha \wedge P_-\tilde f \ .
\end{equation}
At first sight $Q_\alpha$ seems to vanish due to \eqref{eq:vanishint}. 
However, $P_-\tilde f$ need not be in the cokernel of $\iota^*$, and therefore some of the charges $Q_\alpha$ can be non-zero.\footnote{For instance this situation occurs if the flux on the D7-brane and the negative value of the flux on the image-D7-brane can both be written as the pullback of the same two-form in the ambient space. We are indebited to Peter Mayr for drawing our attention to this point and eliminating a misconception in an earlier draft of this paper.}
After eliminating the two-forms $D_\alpha$ in favor of their dual scalars~$\rho_\alpha$ by imposing the self-duality condition on the five-form field-strength of 
$C^{(4)}$,
the Green-Schwarz terms \eqref{eq:GSterms}
modify the local Peccei-Quinn symmetry discussed in \eqref{eq:shift}.
The covariant derivative for $\rho_\alpha$ changes and \eqref{eq:cd1} is 
replaced by
\begin{equation}
   \cov_\mu\rho_\alpha=\partial_\mu\rho_\alpha
    -4\kappa_4^2\mu_7\ell\mathcal{K}_{\alpha bP}\dbbf^b A_\mu
    +4\kappa_4^2\mu_7\ell Q_\alpha A_\mu \ .
\end{equation}
In terms of the chiral coordinates \eqref{eq:G} and \eqref{eq:Tf} 
the contribution proportional to $Q_\alpha$ leave the covariant derivative
of $G^P$ \eqref{eq:Gcov} unchanged while
the fields $T_\alpha$ become charged and a 
covariant derivative of the form
\begin{equation} \label{eq:covT}
   \cov_\mu{T_\alpha}=\partial_\mu T_\alpha+6i\kappa_4^2\mu_7\ell Q_\alpha A_\mu \ 
\end{equation}
is induced.
Thus fluxes $\tilde f$ which lead to non-vanishing $Q_\alpha$
change the gauged isometry \eqref{eq:shift} in that additional 
fields $T_\alpha$ transform non-linearly.

As a consequence of these additional charged chiral fields the D-term is 
also modified. The holomorphic Killing vector field \eqref{eq:Killing1} 
receives an additional contribution from the $T_\alpha$ and reads
\begin{equation} \label{eq:Killing2}
   X=4\kappa_4^2\mu_7\ell \partial_{G^P}-6i\kappa_4^2\mu_7\ell Q_\alpha \partial_{T_\alpha} \ .
\end{equation}
This in turn adjusts the  D-term via 
eq.~\eqref{eq:Dterms} and we find
\begin{equation} \label{eq:fDterm}
   \text{D}\ =\ \frac{12\kappa_4^2\mu_7\ell}{\mathcal{K}}
     \left(\mathcal{K}_{Pa}\dbbf^a-Q_\alpha v^\alpha\right) \
=\ \frac{12\kappa_4^2\mu_7\ell}{\mathcal{K}}\int_{S_P} J\wedge \mathcal{F}\ ,
\end{equation}
where $\mathcal{F}\equiv B-\ell f = \dbbf - \ell\tilde f$.
The corresponding D-term potential is given by
\begin{equation} \label{eq:Vf}
   V_\text{D}=\frac{108\kappa_4^2\mu_7}{\mathcal{K}^2\Real{T_\Lambda}}
     \left(\mathcal{K}_{Pa}\dbbf^a-Q_\alpha v^\alpha\right)^2 \ ,
\end{equation}
where strictly speaking we have to express $v^\alpha$ in terms of the
chiral variables by solving \eqref{eq:Tf}.
We see that whenever there exits at least one odd $(1,1)$-form in the
orientifold one can find a minimum with $\text{D}= 0=V_\text{D}$ by
appropriately choosing $\dbbf$.
On the other hand, if such a form does not exist (i.e.\ $\dbbf\equiv 0$)
and simultaneously $Q_\alpha\neq0$ holds,
 $\text{D}=0$ can only occur for $v^\alpha \to\infty$.
This corresponds to a run-away behavior with no stable minimum.
However, if non-perturbative corrections of the potential stabilize
$v^\alpha$ at some finite value, spontaneous supersymmetry breaking
by this D-term can occur and possibly lead to metastable deSitter
vacua as proposed in \cite{Burgess:2003ic}. 

The potential \eqref{eq:Vf} can receive a further contribution if the 
$U(1)$ gauge symmetry is anomalous. As we already discussed previously
for  $Q_\alpha\neq0$
the fields $T_\alpha$ transform non-linearly as can be seen from 
\eqref{eq:covT}. For $Q_\Lambda\neq0$ this implies that the gauge coupling
$f^\text{D7}$ which is proportional to $T_\Lambda$ transforms and renders
the supergravity action non-invariant.
{}From \eqref{eq:faction} we infer the  anomalous transformation law
\begin{equation} \label{eq:gaugevar}
   \delta\mathcal{S}^{\text{E}}_f=-2\kappa_4^2\mu_7^2\ell Q_\Lambda \int F\wedge F \ .
\end{equation}
This anomaly occurs whenever the charge $Q_\Lambda$ is non-vanishing
and in this case consistency requires the existence of additional charged
chiral fermions.
In order to see their presence let us first discuss the geometric
origin of a non-zero $Q_\Lambda$ in more detail.
{}From \eqref{eq:notvanish} we infer
\begin{equation}\label{QLdef}
   Q_\Lambda=\ell\int_{S^\Lambda}\omega_\Lambda\wedge P_-\tilde f \ ,
\end{equation}
and since $\omega_P=P_-\omega_\Lambda$ it is only consistent
with \eqref{eq:vanishint} if $\omega_P$ is trivial. This, on the other hand, implies that $\omega_\Lambda=\omega_{(1)}+\omega_{(2)}$ with $\omega_{(1)}=\omega_{(2)}$, that is to say
the D7-brane and the image-D7-brane wrap the same homology cycle. By the arguments in the previous section for non-zero $Q_\Lambda$ the form $P_-\tilde f$ can be lifted to a non-trivial form in the ambient space~$Y$ and then the charge $Q_\Lambda$ can be rewritten as
\begin{equation} \label{eq:intersection}
   Q_\Lambda=\ell\int_Y \omega_\Lambda\wedge\omega_\Lambda\wedge P_-\tilde f \ .
\end{equation}
However, the form~$\omega_\Lambda$ is the Poincar\'e dual form of the cycle~$S^\Lambda$ and therefore the non-vanishing integral \eqref{eq:intersection} indicates that the cycle~$S^\Lambda$ has transverse (two-dimensional) self-intersections, which is expressed by the non-triviality of $\omega_\Lambda\wedge\omega_\Lambda$. 

At the self-intersection of the D7-brane there arise additional massless modes from open strings localized at the intersection locus. For both endpoints of these open strings living on the same D7-brane the resulting four-dimensional fields are neutral and thus cannot cancel the anomaly. However, recall that in our notation the cycle $S^\Lambda$ describes the D7-brane wrapped on the internal cycle~$S^{(1)}$ and the image-D7-brane wrapped on $S^{(2)}$ simultaneously. Therefore at a six dimensional locus $\mathcal{I}$ where the D7-brane~$S^{(1)}$ intersects the image-D7-brane~$S^{(2)}$ there do arise additional massless fields which are charged.\footnote{We would like to thank Angel Uranga for explaining this to us.} This is due to the fact that the Kaluza-Klein reduction of the D7-brane $U(1)$ field strength~$F$ has the form 
\begin{equation}
   F(x,y)=F(x)\:P_- + \tilde f(y) + \ldots \ ,
\end{equation}
where $\ldots$ denotes the fluxes $\leftup{Y}{f}$ which are not relevant for this discussion as they arise from pullback forms of the ambient space and hence due to their negative parity they become trivial at the intersection locus $\mathcal{I}$. If we now split this into the $S^{(1)}$ and $S^{(2)}$ contribution one obtains
\begin{align}
\begin{aligned}
   &S^{(1)}:& F(x,y)&=+F(x)+\tilde f^{(1)}(y) \ , \\
   &S^{(2)}:& F(x,y)&=-F(x)+\tilde f^{(2)}(y) \  .
\end{aligned} && 
\tilde f=\tilde f^{(1)}+\tilde f^{(2)} \ ,
\end{align}
Therefore the four-dimensional fields resulting from open strings stretching from $S^{(1)}$ to $S^{(2)}$ have charge $+2$ with respect to the four-dimensional $U(1)$ gauge theory and the number of four-dimensional chiral fermions (or more precisely the asymmetry in the number of chiral fermions) is given by the index $I_{12}$ of the internal Dirac operator at the intersection locus~$\mathcal{I}$ \cite{Blumenhagen:2000wh,Cvetic:2001nr,Cascales:2003pt} 
\begin{equation} \label{eq:index1}
   I_{12}=\int_\mathcal{I} \left(\tilde f^{(1)}-\tilde f^{(2)}\right)
         =\int_\mathcal{I} P_-\tilde f \ .
\end{equation}
As the intersection of the D7-brane with its image can be expressed in terms of the Poincar\'e dual forms $\omega_{(1)}$ and $\omega_{(2)}$ of $S^{(1)}$ and $S^{(2)}$ the integral \eqref{eq:index1} becomes \cite{Cheung:1997az}
\begin{equation}
   I_{12}=\int_Y \omega_{(1)}\wedge\omega_{(2)}\wedge P_-\tilde f
         =\frac{1}{4}\int_Y \omega_\Lambda\wedge\omega_\Lambda\wedge P_-\tilde f 
         =\frac{1}{4}\:Q_\Lambda \ ,
\end{equation}
where we have used that $S^{(1)}$ and $S^{(2)}$ are the same homology
cycles and therefore
$\omega_\Lambda=\omega_{(1)}+\omega_{(2)}=2\omega_{(1)}$. Hence we
find $Q_\Lambda/4$ four-dimensional chiral fermions with charge $+2$,
which then cancel via the four-dimensional analog of the Green-Schwarz
mechanism the variation \eqref{eq:gaugevar}. 

Actually the number of chiral fermions also depends on the type of intersection as some of the fermions might be projected out by the orientifold projection. For this analysis one needs to thoroughly look at the orientifold projection at the intersection locus \cite{Cvetic:2001nr,Cascales:2003pt} which is beyond the scope of this work. However, if tadpole cancellation conditions are fulfilled then by general anomaly flow arguments of intersecting branes all anomalies should cancel \cite{Green:1996dd,Cheung:1997az}.

Let us conclude this section by noting that these additional charged multiplets, which transform linearly with respect to the $U(1)$ gauge theory, also contribute to the D-term \eqref{eq:fDterm} according to 
\begin{equation} \label{eq:DtermShift}
   \text{D}\rightarrow\text{D}+2\kappa_4^2\mu_7\ell\sum_i q_i\abs{X_i}^2 \ ,
\end{equation}
where $X_i$ denotes chiral fields with charge $q_i$ arising from brane
intersections. However these additional terms have not been calculated
in the Kaluza-Klein reduction described in the previous sections but
can only be inferred via the anomaly analysis.

Before we turn to the computation of the D- and F-terms from
fermionic couplings let us discuss the origin of the difficulty  in
computing
the scalar potential directly from a Kaluza-Klein analysis. The reason is that in the derivation of the effective action one wants to preserve $\mathcal{N}=1$ supersymmetry or in other words wrap a D7-brane which satisfies a BPS-condition on a supersymmetry four-cycle. For a space-time filling D7-brane the BPS-calibration condition for the internal four cycle $S^\Lambda$ was derived in ref.~\cite{Marino:1999af} to be
\begin{equation} \label{eq:BPS1}
   \dd^4\xi\sqrt{\det(\hat g + \dbbf^a\:\iota^*\omega_a-\ell\tilde f)}
      =\frac{1}{2}e^{-i\theta}\left(J+i\dbbf^a\:\iota^*\omega_a-i\ell\tilde f\right)\wedge
       \left(J+i\dbbf^b\:\iota^*\omega_b-i\ell\tilde f \right) \ ,
\end{equation}
where $\hat g$ is the Calabi-Yau metric in the string frame. The real
constant $\theta$ parametrizes the linear combination of supersymmetry
parameters preserved by the BPS D7-brane. Therefore in order to
maintain $\mathcal{N}=1$ supersymmetry in Calabi-Yau orientifolds the
parameter $\theta$ must be in accord with the supersymmetry truncation
resulting from the orientifold projection \eqref{eq:proj}, which fixes
$\theta=0$ \cite{Jockers:2004yj}. As a consequence since the left hand
side of eq.~\eqref{eq:BPS1} is manifestly real the right hand side
also needs to be real with $\theta=0$. In other words the imaginary
part of \eqref{eq:BPS1} vanishes in the case of unbroken supersymmetry
which yields the condition 
$\mathcal{F}\wedge J = (\dbbf-\ell\tilde f)\wedge J = 0$. Integrated over $S^P$ and using \eqref{eq:K} and \eqref{eq:notvanish} one obtains 
\begin{equation} \label{eq:BPS2}
 \int_{S_P} J\wedge \mathcal{F} = \mathcal{K}_{Pa}\dbbf^a-Q_\alpha
 v^\alpha\ =\ 0 \ ,
\end{equation}
which precisely corresponds to the condition that the D-term given in \eqref{eq:fDterm} vanishes. In this case \eqref{eq:BPS1} reduces to
\cite{Marino:1999af,Jockers:2004yj} 
\begin{equation} \label{eq:BPS3}
   \dd^4\xi\sqrt{\det(\hat g + \dbbf^a\:\iota^*\omega_a-\ell\tilde f)}
      =\frac{1}{2}J\wedge J-\frac{1}{2}\left(\dbbf^a\:\iota^*\omega_a-\ell\tilde f\right)\wedge
       \left(\dbbf^b\:\iota^*\omega_a-\ell\tilde f \right) \ .
\end{equation}

The deviation from the BPS condition is a measure of supersymmetry breaking and is reflected in the appearance of the D-term \eqref{eq:fDterm}. Note that the expression \eqref{eq:BPS2} enters in the D-term linearly whereas it appears quadratically in the scalar potential.
Apart from the D-term potential $V_\text{D}$ given in \eqref{eq:Vf}
the fluxes $\tilde f$ also induce a superpotential or in other words
an F-term potential $V_\text{F}$.
We were unable to compute these potential terms, which are quadratic in the fluxes, from
a Kaluza-Klein reduction of the bosonic terms.
However, in order to determine the scalar potential of the low energy
supergravity action we need not just rely on the bosonic side of the
theory, but instead we can also compute parts of the fermionic
supergravity action in order to gain via supersymmetry information
about the scalar potential for the bosons. Therefore we turn to the
analysis of the fermions in the next section.


\section{Fermionic spectrum and couplings  of D7-branes in Calabi-Yau orientifolds} 
\label{sec:fermions}


As we have argued in the previous section it is difficult to compute the
scalar potential from the bosonic terms of a Kaluza-Klein reduction.
However, both the superpotential and the D-terms can be computed more
reliably from the fermionic couplings of a Kaluza-Klein reduced theory.
The reason is that in certain fermionic couplings the D-terms
and the F-terms (i.e.\ derivatives of the superpotential) appear
linearly and thus can be obtained by a first order perturbation theory
around the background. 

In this section we determine the fermionic spectrum of the D7-brane excitations and compute some of their scalar field dependent couplings to the gravitino.
The fermionic D7-brane fields appear in the super Dirac-Born-Infeld action and the super Chern-Simons action of a super-D7-brane which we introduce in section~\ref{sec:DBaction}. From these superspace worldvolume actions the fermionic
excitations are determined in section~\ref{sec:fspec}
and shown to complete the $\mathcal{N}=1$ 
supersymmetric multiplets of section~\ref{sec:review}.
 In section~\ref{sec:fsugra} we review and set our notation for the fermionic part of the $\mathcal{N}=1$ supergravity action in four dimensions. This is also the preparation for section~\ref{sec:fkin} in which the kinetic terms of the fermions are derived from the super Dirac-Born-Infeld action. Then in section \ref{sec:DFterms} the super Chern-Simons action serves as the starting point to deduce particular fermionic couplings from which we read off the D-terms and the flux induced superpotential of the four-dimensional effective supergravity theory. Finally in section~\ref{sec:HCS} we show how this superpotential can also be obtained by dimensional reduction of the holomorphic Chern-Simons action of ref.~\cite{Witten:1992fb}.


\subsection{Super-D7-brane action} \label{sec:DBaction}


In the large radius limit of Calabi-Yau compactifications a D$p$-brane can be viewed geometrically as the embedding of its $p+1$-dimensional worldvolume~$\mathcal{W}$ in the ten-dimensional target space manifold $M^{9,1}$, namely $\varphi:\mathcal{W}\hookrightarrow M^{9,1}$ with the embedding map $\varphi$. In order to generalize this concept to a super-D$p$-brane (i.e.\ we also want to capture the fermionic degrees of freedom) the worldvolume $\mathcal{W}$ needs to be embedded in the target-space supermanifold of the considered string theory.

Here we focus on D7-branes which arise as extended objects in type~IIB string theory. In superspace the corresponding type~IIB supergravity theory is formulated on the supermanifold $M^{9,1|2}$ with ten even dimensions and two odd dimensions. Locally this supermanifold is described by the superspace coordinates $Z^{\check M}=(x^M,\pair{\theta})$ with the ten bosonic coordinates $x^M$ and the pair of fermionic coordinates $\pair{\theta}=\left(\theta^1,\theta^2\right)$. As type~IIB string theory is chiral with $\mathcal{N}=2$ supersymmetry in ten dimensions the pair of fermionic coordinates $\pair{\theta}$ consists of two Majorana-Weyl spinor of $SO(9,1)$ with the same chirality. Hence $\theta^1$ and $\theta^2$ are related by the $SO(2)$ R-symmetry of type~IIB supergravity \cite{Howe:1983sr}.

Now in this formulation the super-D7-brane appears as the embedding of the eight dimensional worldvolume $\mathcal{W}$ in the supermanifold $M^{9,1|2}$. The embedding $\super{\varphi}:\mathcal{W}\hookrightarrow M^{9,1|2}$ is now described by the supermap $\super{\varphi}$ which maps a point in the worldvolume~$\mathcal{W}$ to a superpoint in the target space supermanifold $M^{9,1|2}$.

The super Dirac-Born-Infeld action for a single super-D7-brane becomes in the string frame \cite{Cederwall:1996,Bergshoeff:1996tu}
\begin{equation} \label{eq:SDBI}
   \mathcal{S}^\text{sf}_\text{DBI}=-\mu_7 \int_\mathcal{W}\dd^8\xi e^{-\super{\varphi^*\phi}}
       \sqrt{-\det\left(\super{\varphi^*}\left(\super{g_{10}}+\super{B}\right)_{ab}-\ell F_{ab} \right)} \ ,
\end{equation}
whereas the super Chern-Simons action reads
\begin{equation} \label{eq:SCS}
   \mathcal{S}_\text{CS}=\mu_7 \int_\mathcal{W}\sum_q\super{\varphi^*}\left(\super{C^{(q)}}\right)
       e^{\ell F-\super{\varphi^*B}} \ .
\end{equation}
Both actions resemble their bosonic analogs but the bulk fields $\super{g_{10}}$, $\super{B}$, $\super{\phi}$ and $\super{C^{(q)}}$ have been promoted to bulk superfields with their lowest components being the corresponding bosonic fields. These superfields are then pulled back with the supermap~$\super{\varphi}$.\footnote{Note that the 
pulled-back quantities $\super{\varphi}^*(\cdot)$ contain no odd components because $\super{\varphi}$ is a map from an ordinary manifold into a supermanifold. As a consequence the integrals in \eqref{eq:SDBI} and \eqref{eq:SCS} are integrals only over bosonic coordinates.}
The gauge field strength $F$ on the brane, however, remains a bosonic object.

The super-D7-brane possesses also a local fermionic gauge symmetry called $\kappa$-sym\-metry which removes half of its fermionic degrees of freedom \cite{Bergshoeff:1996tu,Bergshoeff:19972000}.  In Calabi-Yau 
orientifold compactifications  the supersymmetry is reduced from $\mathcal{N}=2$ to $\mathcal{N}=1$ by the orientifold projection~$\mathcal{O}$. A BPS D7-branes respects the same linear combination of supercharges which is invariant under the orientifold projection~$\mathcal{O}$. Thus if we also truncate the effective theory of the open string sector with the orientifold projection~$\mathcal{O}$, $\kappa$-symmetry is simultaneously fixed and the correct number of fermionic degrees of freedom is obtained. 


\subsection{The fermionic D7-brane spectrum} \label{sec:fspec}


Before we enter the discussion of the fermionic D7-brane spectrum, let us pause and discuss the gravitinos of the bulk theory. This also allows us to recall some techniques which are important for the remainder of this section.\footnote{%
Our spinor conventions are assembled in appendix~\ref{sec:conv}.}

In compactifying string theory on a six dimensional manifold the structure group $SO(9,1)$ of $M^{9,1}$ reduces to $SO(3,1)\times SO(6)$ due to the product structure \eqref{eq:Pansatz}. Compactifications on a six dimensional complex manifolds reduce the structure group $SO(6)$ further to $U(3)\cong SU(3)\times U(1)$ so that we have
\begin{equation} \label{eq:Decomp3C}
   SO(9,1)\rightarrow SO(3,1)\times SO(6)\rightarrow SO(3,1)\times SU(3)\times U(1) \ .
\end{equation}
Correspondingly the  Weyl spinor $ \spinrep{16'}$ of $SO(9,1)$ decomposes into 
representations of $SO(3,1)\times SO(6)$ or 
$SO(3,1)\times SU(3)\times U(1)$ respectively
\begin{align} \label{eq:D10_SU3}
   \spinrep{16'}\rightarrow (\spinrep{2},\spinrep{\bar 4})\oplus (\spinrep{\bar 2},\spinrep{4}) 
\rightarrow (\spinrep{2},\spinrep{\bar 3_1})\oplus(\spinrep{2},\spinrep{\bar 1_{-3}})
\oplus (\spinrep{\bar 2},\spinrep{3_{-1}})\oplus (\spinrep{\bar 2},\spinrep{1_3})\ .
\end{align}
$\spinrep{2},\spinrep{\bar 2}$ are the two Weyl spinors of $SO(3,1)$, $\spinrep{4}, \spinrep{\bar 4}$ are the two Weyl spinors of $SO(6)$,
 $\spinrep{3}, \spinrep{\bar 3}$ are the fundamentals of $SU(3)$
and  $\spinrep{1}, \spinrep{\bar 1}$ are $SU(3)$ singlets 
(the subscript denotes their $U(1)$ charge).

For complex threefolds the Clifford algebra for the $SO(6)$ Dirac gamma matrices~$\check\gamma^m$ can be rewritten in terms of complex coordinate indices which then obey\footnote{The $\:\check{}\:$ denotes six-dimensional bulk quantities.}
\begin{align}\label{eq:gammaalgebra}
   \ac{\check\gamma^i}{\check\gamma^{\bar\jmath}}=2 g^{i\bar\jmath} \ , &&
   \ac{\check\gamma^i}{\check\gamma^j}=0 \ , &&
   \ac{\check\gamma^{\bar\imath}}{\check\gamma^{\bar\jmath}}=0 \ , &&
   \ac{\check\gamma^i}{\check\gamma}=\ac{\check\gamma^{\bar\jmath}}{\check\gamma}=0 \ ,
\end{align} 
where $\check\gamma$ is the six-dimensional analog of $\gamma^5$.
These relations allow us to interpret the Dirac gamma-matrices with holomorphic indices as raising and lowering operator acting on some `ground state'~$\check\singspin$ and its `conjugate ground state'~$\check\singspin^\dagger$ 
\begin{align} \label{eq:checksingspin}
   \check\gamma^i\check\singspin=0 \ , && 
   \check\singspin^\dagger\check\gamma^{\bar\imath}=0 \ .
\end{align}
$\check\singspin, \check\singspin^\dagger$ are the singlets 
$\spinrep{1}, \spinrep{\bar 1}$
of $SU(3)$ which obey the chirality property 
\begin{align}\label{eq:xichiral}
   \check\gamma\check\singspin=+\check\singspin \ , 
   && \check\singspin^\dagger\check\gamma=-\check\singspin^\dagger \ .
\end{align}
Note that the conditions \eqref{eq:checksingspin} are maintained on the whole complex manifold~$Y$ because the structure group $U(3)\cong SU(3)\times U(1)$ of complex threefolds does not mix the gamma matrices $\check\gamma^i$ with the gamma matrices $\check\gamma^{\bar\jmath}$. However, for a generic complex manifolds the $SU(3)$ singlets $\singspin$ and $\singspin^\dagger$ transform under the $U(1)$ part of \eqref{eq:Decomp3C}, which is the $U(1)$ part in the spin-connection of the internal space. Thus for a complex manifold we have two `charged' spinor singlets $\singspin\otimes\mathcal{L}^*$ and $\singspin^ \dagger\otimes\mathcal{L}$ which are tensors of the $SU(3)\times U(1)$ bundle. Here $\mathcal{L}$ and $\mathcal{L}^*$ are sections of the line bundle corresponding to the $U(1)$ in \eqref{eq:Decomp3C}. For Calabi-Yau manifolds the first Chern class of the tangent bundle vanishes. This also implies that the line-bundle arising from the $U(1)$ part of the spin connection is trivial, and as a consequence one finds for a Calabi-Yau manifold two globally defined singlets $\singspin$ and $\singspin^\dagger$ which are in addition covariantly constant. 

In type~IIB string theory there are two ten-dimensional Majorana-Weyl gravitinos~$\pair{\Psi}_{M}=(\Psi_M^1,\Psi_M^2)$, which have the same chirality 
\begin{equation}\label{eq:Psichiral}
   \Gamma\pair{\Psi}_M = - \pair{\Psi}_M \ .
\end{equation}
Due to \eqref{eq:D10_SU3} the ten-dimensional gravitinos~$\pair{\Psi}_M$ compactified on the Calabi-Yau manifold~$Y$ give rise to a set of two massless four-dimensional Weyl gravitinos $\pair{\psi}_\mu(x)$\footnote{%
Since $\pair{\Psi}_M$ are Majorana-Weyl gravitinos, one can choose a Majorana basis such that $\pair{\Psi}_M^*=\pair{\Psi}_M$. This condition implies that the decomposed spinors in  \eqref{eq:D10_SU3} are complex conjugate to each other and as a consequence also the spinors
in the Kaluza-Klein Ansatz \eqref{eq:4Dgrav} are complex conjugates.\label{ft:MJ}} 
\begin{equation} \label{eq:4Dgrav}
   \pair{\Psi}_\mu
     =\pair{\bar\psi}_\mu(x)\otimes\check\singspin(y)
       +\pair{\psi}_\mu(x)\otimes\check\singspin^\dagger(y)\ .
\end{equation}
Using \eqref{eq:Psichiral}, \eqref{eq:gammadecomp} and \eqref{eq:xichiral}
the four-dimensional gravitinos have to obey
$\hat\gamma^5\pair{\psi}_\mu=+\pair{\psi}_\mu$ and $\hat\gamma^5\pair{\bar\psi}_\mu=-\pair{\bar\psi}_\mu$. 

In Calabi-Yau orientifold compactifications only one linear combination of the two four-dimensional spinors in \eqref{eq:4Dgrav} is invariant with respect to the orientifold projection \eqref{eq:proj}. Recall that the two ten-dimensional gravitinos~$\Psi_M$ of the type~IIB superstring theory arise from the R-NS and NS-R sector respectively. The worldsheet operator $(-1)^{F_L}\Omega_p$ maps the R-NS sector to the NS-R sector and vice versa and adds an additional minus sign for the NS-R sector \cite{Vafa:1995gm,Dabholkar:1996pc}. Thus for $\Psi_M^1$ from R-NS and $\Psi_M^2$ from the NS-R sector one obtains in terms of the Pauli matrices $\check\sigma^1,\check\sigma^2,\check\sigma^3$ for the two component spinor $\pair{\Psi}_M$ the transformation behavior
\begin{equation} \label{eq:inv1}
   (-1)^{F_L}\Omega_p \pair{\Psi}_M = -i\check\sigma^2\:\pair{\Psi}_M \ .
\end{equation}
As a consequence the operator $(-1)^{F_L}\Omega_p$ acts in the same way on the Kaluza-Klein modes of $\pair{\Psi}_\mu$ given in eq.~\eqref{eq:4Dgrav}.

The next step is to determine the action of the pullback of the holomorphic involution~$\sigma^*$ on the spinors~$\check\singspin$ and $\check\singspin^\dagger$. Due to eq.~\eqref{eq:invOmega} the unique holomorphic three-form is odd with respect to the involution~$\sigma$. This implies that at each point in the tangent space of $Y$ the involution~$\sigma$ generates a rotation by $\pm\pi$. Since the spinor singlets~$\singspin$ and $\singspin^\dagger$ are sections of the spin bundle of $Y$ the rotation by~$\pm\pi$ lifts in the spin bundle to a phase factor of $e^{\pm i\pi/2}$. This implies the transformation properties $\sigma^*\check\singspin=\pm i\check\singspin$ and $\sigma^*\check\singspin^\dagger=\mp i\check\singspin^\dagger$.\footnote{For toroidal models the factor $i$ follows directly from the R-NS and NS-R gravitino vertex operator: A four-dimensional orientifold theory with O7~planes is obtained by first compactifying type~I string theory on a six torus and then T-dualizing two distinct internal directions. In this case the orientifold projection becomes $(-1)^{F_L}R_{89} \Omega_p$ where $R_{89}$ is the reflection operator along the T-dualized directions \cite{Dabholkar:1996pc}. Note that such a reflection corresponds to a rotation by $\pm\pi$ and thus produces for worldsheet spinors in the R-sector a phase factor $e^{\pm i\pi/2}=\pm i$ which also appears for the gravitinos in the R-NS and NS-R sector.}
Applied to the four-dimensional gravitinos and using \eqref{eq:inv1} 
we obtain 
\begin{align} \label{eq:inv2}
   \mathcal{O}\pair{\psi}_\mu\otimes\check\singspin^\dagger
      = \check\sigma^2\: \pair{\psi}_\mu\otimes\check\singspin^\dagger \ , &&
   \mathcal{O}\pair{\bar\psi}_\mu\otimes\check\singspin
      = \pair{\bar\psi}_\mu\otimes\check\singspin\: \check\sigma^2 \ .
\end{align}
Thus the four-dimensional Weyl gravitinos $\psi_\mu$ and $\bar\psi_\mu$ of the $\mathcal{N}=1$ low energy effective orientifold theory are given as the 
invariant linear combinations 
\begin{align} \label{eq:gravitinos}
   \psi_\mu\otimes\check\singspin^\dagger = \tfrac{1}{2}\left(\id+\check\sigma^2\right)\:
       \pair{\psi}_\mu\otimes\check\singspin^\dagger \ , &&
   \bar\psi_\mu\otimes\check\singspin = 
       \pair{\bar\psi}_\mu\otimes\check\singspin\: 
       \tfrac{1}{2}\left(\id+\check\sigma^2\right) \ .
\end{align}

After this interlude on the gravitinos we turn to the fermionic matter fields. For the fields localized on the D7-brane the structure group $SO(6)$ of the tangent bundle of the internal six-dimensional space splits into $SO(4)\times SO(2)$. Here $SO(4)$ is the structure group of the tangent bundle of the four-dimensional internal D7-brane cycle $S^\Lambda$, whereas $SO(2)$ is the structure group of the two-dimensional normal bundle of $S^\Lambda$. Moreover for the D7-branes under consideration we always assume that the pullback tangent bundle $\iota^*\tbundle{Y}$ splits holomorphically into the direct sum $\tbundle{S^\Lambda}\oplus\nbundle{S^\Lambda}$, i.e. the structure group of the tangent and normal bundle reduces to $U(2)\times U(1)$. Hence we have the following chain of subgroups 
\begin{multline} \label{eq:gr}
   SO(3,1)\times SO(6)\xrightarrow{\ \iota^*\tbundle{Y}\rightarrow\tbundle{S^\Lambda}\oplus\nbundle{S^\Lambda}\ } 
      SO(3,1)\times SO(4) \times SO(2) \\
      \xrightarrow{\ \text{holomorphicity}\ } SO(3,1)\times\left(SU(2)\times U(1)\right)\times U(1) \ ,
\end{multline}
which tells us that the D7-brane worldvolume fields are appropriate representations of $SO(3,1)\times SU(2)\times U(1)\times U(1)$. 

Analogously to \eqref{eq:D10_SU3} the ten-dimensional spinor representation
decomposes under $SO(3,1)\times SO(4)\times SO(2)$ according to
\begin{equation} \label{eq:D10_SO4}
   \spinrep{16'}\rightarrow \left(\spinrep{2},\spinrep{2},\spinrep{\bar 1}\right)\oplus
   \left(\spinrep{2},\spinrep{2'},\spinrep{1}\right)\oplus
   \left(\spinrep{\bar 2},\spinrep{2},\spinrep{1}\right)\oplus
   \left(\spinrep{\bar 2},\spinrep{2'},\spinrep{\bar 1}\right) \ ,
\end{equation}
with the two Weyl spinors $\spinrep{2}$ and $\spinrep{2'}$ of $SO(4)$ and  the
$SO(2)$ complex conjugated singlets
$\spinrep{1}, \spinrep{\bar 1}$.

As before the ten-dimensional Dirac gamma matrices~$\Gamma^M$ are decomposed into a tensor product of Dirac gamma matrices of the subgroups of \eqref{eq:gr} according to \eqref{eq:DG}. Then the Dirac gamma matrices~$\gamma^m$ of the tangent bundle structure group $SO(4)$ are combined into holomorphic and anti-holomorphic gamma matrices which fulfill the anti-commutation relations
\begin{align}
   \ac{\gamma^i}{\gamma^{\bar\jmath}}=2 g^{i\bar\jmath} \ , &&
   \ac{\gamma^i}{\gamma^j}=0 \ , &&
   \ac{\gamma^{\bar\imath}}{\gamma^{\bar\jmath}}=0 \ , &&
   \ac{\gamma^i}{\gamma}=\ac{\gamma^{\bar\jmath}}{\gamma}=0 \ .
\end{align} 
Similarly as in the case of the Calabi-Yau threefold the holomorphic $SO(4)$ gamma matrices are interpreted as raising and lowering operators which are used to construct the `ground states' $\singspin$ and the `conjugate ground state' $\singspin^\dagger$ 
\begin{align} \label{eq:singspin}
   \gamma^i\singspin=0 \ , && \singspin^\dagger\gamma^{\bar\imath}=0 \ .
\end{align}
These `ground states' correspond to $SU(2)$ spinor singlets of the same chirality\footnote{The singlets $\singspin$ and $\singspin^\dagger$ have the same chirality for structure groups $SU(2k)$ and different chiralities for structure groups $SU(2k+1)$ because the `conjugate ground states' $\singspin^\dagger$ is also obtained by acting with all raising operators on the `ground state' $\singspin$. Thus for the group $SU(2k)$ there is an even number of raising operators which yields for $\singspin$ and $\singspin^\dagger$ the same chirality, whereas for $SU(2k+1)$ the odd number of raising operators results in different chiralities for $\singspin$ and $\singspin^\dagger$.}
\begin{align} \label{eq:G5Ch}
   \gamma\singspin=+\singspin \ , && \singspin^\dagger\gamma=+\singspin^\dagger \ .
\end{align}

Analogously as in the case of the ambient Calabi-Yau threefold~$Y$ the internal D7-brane cycle~$S^\Lambda$ has two spinor singlets $\singspin\otimes\mathcal{L^*}$ and $\singspin^\dagger\otimes\mathcal{L}$ which are globally defined tensors of the $SU(2)\times U(1)$ bundle. The sections $\mathcal{L}^*$ and $\mathcal{L}$ are again associated to the $U(1)$ part of the spin connection. In general these line bundles are non-trivial as $S^\Lambda$ need not be a Calabi-Yau manifold with trivial first Chern class. However, the spinors of $S^\Lambda$ relevant for our analysis transform under the induced spin connection of the ambient space~$Y$ which is a connection with respect to both the tangent and the normal bundle of $S^\Lambda$. Therefore due to the triviality of the $U(1)$ part of the spin connection in the ambient space~$Y$ the overall $U(1)$ `charge' of the induced spinors must also be trivial and as a consequence the line bundles associated to the structure group $U(1)$ of the normal bundle~$\nbundle{S^\Lambda}$ must be dual to the line bundles of the $U(1)$ part in the structure group of the tangent bundle~$\tbundle{S^\Lambda}$. Hence the two induced spinor singlets are globally defined tensors of $SU(2)\times U(1)\times U(1)$ and take the form
\begin{align} 
   (\singspin\otimes\mathcal{L}^*)\otimes\mathcal{L} \cong \singspin\ , &&
   (\singspin^\dagger\otimes\mathcal{L})\otimes\mathcal{L}^* \cong \singspin^\dagger\ ,
\end{align}
which behave like `neutral' spinors with respect to the $U(1)$ part of the induced spin connection.\footnote{Mathematically this neutrality is a consequence of the Whitney formula for the first Chern class. Since $\iota^*\tbundle{Y}\cong\tbundle{S^\Lambda}\oplus\nbundle{S^\Lambda}$ holomorphically, the Whitney formula applied to the Calabi-Yau manifold~$Y$ yields $0=\iota^*c_1(\tbundle{Y})=c_1(\tbundle{S^\Lambda})+c_1(\nbundle{S^\Lambda})$. Thus the line bundle associated to the tangent bundle must be dual to the line bundle associated to the normal bundle.} Note that the chiralities of the line bundles $\mathcal{L}$ and $\mathcal{L}^*$ viewed as spinors of the normal bundle $\nbundle{S^\Lambda}$ are given by 
\begin{align} \label{eq:G3Ch}
   \tilde\gamma \mathcal{L} = +\mathcal{L} \ , && \tilde\gamma \mathcal{L}^* = -\mathcal{L}^* \ ,
\end{align}
with the Dirac gamma matrices of $SO(2)$ as introduced in appendix~\ref{sec:Spinors}.

As in the bosonic part of the D7-brane the dynamics of the super-brane is captured by fluctuations of the embedding supermap $\super{\varphi}$. 
These fluctuations determine the supersymmetric spectrum including 
the bosonic excitations and their fermionic superpartners which are the focus of this section. In order to derive from \eqref{eq:SDBI} and \eqref{eq:SCS} the low energy effective action for the fermionic excitations one needs to compute the pullback of the superfields with respect to the fluctuating supermap $\super{\varphi}$, which takes for the fermionic fluctuations the simple form \cite{Grisaru:1997ub,Marolf:2003vf}
\begin{equation}
   \super{\varphi}:\mathcal{W}\hookrightarrow M^{9,1|2},\qquad
      \xi\mapsto\left(\varphi(\xi),\pair{\Theta}(\xi)\right) \ .
\end{equation}
Hence the pullback with respect to this supermap simply amount to replacing the superspace coordinates~$\pair\theta$ by $\pair{\Theta}(\xi)$ \cite{Grana:2002,Millar:2000ib}. Due to the dependence on $\xi$ the fermionic fields~$\pair\Theta(\xi)$ are localized on the worldvolume~$\mathcal{W}$ of the D7-brane and contain all the fermionic degrees of freedom of the super-D7-brane. 
Note that $\pair\Theta(\xi)$ has the same transformation behavior as the odd supercoordinates~$\pair\theta$, that is to say they are Majorana-Weyl fermions transforming in the Weyl representation $\spinrep{16'}$ of $SO(9,1)$.

The next task is to determine the massless Kaluza-Klein modes resulting from the Majorana-Weyl spinors $\pair\Theta(\xi)$ compactified on the worldvolume $\mathbb{R}^{3,1}\times S^\Lambda$.  As for the gravitinos only one linear combination of $\pair\Theta(\xi)$ is invariant with respect to the orientifold projection~$\mathcal{O}$. The fermions~$\pair\Theta(\xi)$ are the fluctuations of the
odd superspace coordinates~$\pair\theta$, which in turn correspond to the infinitesimal supersymmetry parameters for supersymmetry variations. Since
the gravitinos are the gauge fields for local supersymmetry the supersymmetry parameters $\pair\theta$ and their fluctuations $\pair\Theta$ transform
exactly like the gravitinos under $\mathcal{O}$. 
Thus the projector $\tfrac{1}{2}\left(\id+\mathcal{O}\right)$ acting on $\pair\Theta$ becomes $\tfrac{1}{2}\left(\id+\check\sigma^2\right)$ as in \eqref{eq:inv2}, and we define
\begin{equation} \label{eq:inv3}
   \Theta(\xi)=\tfrac{1}{2}(\id+\mathcal{O})\:\pair\Theta(\xi) \ .
\end{equation} 
Then the projected ten-dimensional Majorana-Weyl spinor~$\Theta(\xi)$
transforming as $\spinrep{16'}$ of $SO(9,1)$ needs to be decomposed
into representations of the subgroups in eq.~\eqref{eq:gr} according
to \eqref{eq:D10_SO4}. 

Since $\singspin$ and $\singspin^\dagger$ are constant sections on
$S^\Lambda$ (or zero-forms) we can identify
$\gamma^{\bar\imath}\singspin$ and $\singspin^\dagger\gamma^i$ with
$(0,1)$-forms and $(1,0)$-forms and
$\gamma^{\bar\imath}\gamma^{\bar\jmath}\singspin$ and
$\singspin^\dagger\gamma^i\gamma^j$ with $(0,2)$-forms and $(2,0)$-forms. 
Furthermore, we need to expand $\Theta(\xi)$ into 
fermionic modes which are invariant under the orientifold projection $\mathcal{O}$.
By supersymmetry we already know that the invariant fermionic modes are only identified with negative parity forms in order to match the negative parity of their bosonic superpartners.
Finally, we only keep massless fermionic excitations, which are
zero-modes of the internal Dirac operator. As explained in the
previous paragraph the relevant Dirac operator is induced from the
ambient Calabi-Yau space~$Y$ for which the $U(1)$ part of the spin
connection is trivial, and in this case the square of the Dirac
operator can be identified with the Laplace operator. This implies
that the massless fermionic excitations are in one-to-one
correspondence with the odd  harmonic $(p,q)$-forms. 
Using \eqref{eq:G5Ch} and \eqref{eq:G3Ch} this leads to the Kaluza-Klein expansion
\begin{equation} \label{eq:decomptheta}
\begin{split}
   \Theta(\xi)\ =\  &\
       N_\lambda\:\lambda(x)\otimes P_-\singspin^\dagger
          +\bar N_{\bar\lambda}\:\bar\lambda(x)\otimes
           P_-\singspin\\
& +N_{\chi_I}\:\chi_I(x)\otimes A^I_{\bar\imath}\gamma^{\bar\imath}\singspin
          +\bar N_{\bar\chi_{\bar I}}\:\bar\chi_{\bar I}(x)
           \otimes\bar A^{\bar I}_i\singspin^\dagger\gamma^i   \\
     &+N_{\chi^A}\:\chi^A(x)\otimes\tfrac{1}{2} \tilde s_{A\: ij} 
          \singspin^\dagger\gamma^j\gamma^i
          +\bar N_{\bar\chi^{\bar A}}\:\bar\chi^{\bar A}(x)\otimes
           \tfrac{1}{2}\tilde s_{\bar A\:\bar\imath\bar\jmath} 
          \gamma^{\bar\imath}\gamma^{\bar\jmath}\singspin\ ,
\end{split}
\end{equation}
where $\lambda(x), \chi_I(x), \chi^A(x)$ are four-dimensional Weyl spinors. $P_-$ is the harmonic zero form of $S^\Lambda$ defined in \eqref{eq:P}. $A^I$ is a basis of odd (0,1)-forms on $S^\Lambda$ while $\tilde s_A$ is a bases of odd (2,0)-forms, both of which we already introduced in Table~\ref{tab:spec}. For the moment we also included a set of normalization constants $N_\lambda$, $N_{\chi^A}$ and $N_{\chi_I}$ which are determined in the next section.  Note that the Majorana property of the spinor $\Theta(\xi)$ implies that the decomposition \eqref{eq:D10_SO4} must fulfill a reality condition. This is reflected in the expansion \eqref{eq:decomptheta} in that for each term the complex conjugate term also appears.

Thus altogether we conclude that the four-dimensional massless fermionic modes invariant under $\mathcal{O}$ are identified with the negative harmonic forms of the cycle~$S^\Lambda$, which justifies the expansion of \eqref{eq:decomptheta} into the forms $P_-$, $\tilde s_A$ and $A^I$. The resulting massless fermionic spectrum from \eqref{eq:decomptheta} is summarized in Table~\ref{tab:D7spec}, which also illustrates the formation of four-dimensional $\mathcal{N}=1$ multiplets by combining the fermions with their bosonic superpartners of Table~\ref{tab:spec}.

\begin{table}
\begin{center}
\begin{tabular}{|c|c|c|c|}
   \hline
      \bf bos. fields  &  \bf ferm. fields  &  \bf multiplet & \bf multiplicity
      \rule[-1.5ex]{0pt}{4.5ex} \\
   \hline
   \hline
      $A_\mu$  &  $\lambda$, $\bar\lambda$  &  vector  &  $1$
      \rule[-1.5ex]{0pt}{4.5ex} \\
   \hline
      $\db^A$, $\bar\db^{\bar A}$  &  $\chi^A$, $\bar\chi^{\bar A}$  &  
      chiral  &  $\dim H_{\bar\partial,-}^{(2,0)}(S^\Lambda)$ 
      \rule[-1.5ex]{0pt}{4.5ex} \\
   \hline
      $a_I$, $\bar a_{\bar I}$  &  $\chi_I$, $\bar\chi_{\bar I}$  &  
      chiral  &  $\dim H_{\bar\partial,-}^{(0,1)}(S^\Lambda)$ 
      \rule[-1.5ex]{0pt}{4.5ex} \\
   \hline
\end{tabular} 
\caption{D7-brane spectrum in four dimensions and $\mathcal{N}=1$ multiplets} \label{tab:D7spec} 
\end{center}
\end{table}


\subsection{Fermionic terms of $D=4$, $\mathcal{N}=1$ supergravity action} \label{sec:fsugra}


In section~\ref{sec:review} we have introduced the bosonic part of the
four-dimensional $\mathcal{N}=1$ supergravity action for generic
scalar fields $M^M$ in chiral multiplets and vectors $V^\Gamma$ in
vector multiplets. Their supersymmetric partners are Weyl fermions
$\chi^M$ and $\lambda^\Gamma$ which complete the chiral multiplets
$(M^M,\chi^M)$ and the vector multiplets~$(V^\Gamma,\lambda^\Gamma)$. 
In addition the gravitational multiplet contains the four-dimensional
metric $\eta_{\mu\nu}$ and the gravitino $\psi_\mu$.
 The complete $\mathcal{N}=1$ supergravity action for these multiplets
 is given by
\begin{equation}
   \mathcal{S}^{\mathcal{N}=1}_\text{SUGRA}=\mathcal{S}_\text{Bosons}+\mathcal{S}_\text{Fermions}
      +\mathcal{S}_\text{Couplings} \ ,
\end{equation}
where the bosonic part $\mathcal{S}_\text{Bosons}$ is specified in \eqref{eq:4Dbos}, and where $\mathcal{S}_\text{Fermions}$ contains the kinetic terms of the fermions and $\mathcal{S}_\text{Couplings}$ are the fermionic interaction terms. In the conventions of ref.~\cite{Wess:1992} the kinetic terms of the fermions in the supergravity action read
\begin{equation} \label{eq:4Dferm}
\begin{split}
    \mathcal{S}_\text{Fermions}=& - \frac{1}{\kappa_4^2}\int\dd^4x\:\sqrt{-\eta} 
     \left[\vphantom{\frac{1}{2}}
     \left(-\epsilon^{\mu\nu\rho\tau}\bar\psi_\mu\bar\sigma_\nu\nabla_\rho\psi_\tau +
     i K_{M\bar N}\bar\chi^{\bar N}\bar\sigma^\mu\nabla_\mu\chi^M \right)\right. \\
    &+\left.\frac{i}{2} (\Real f)_{\Gamma\Delta}\left(\lambda^\Gamma\sigma^\mu\nabla_\mu\bar\lambda^\Delta+ 
     \bar\lambda^\Gamma\bar\sigma^\mu\nabla_\mu\lambda^\Delta\right)
     -\frac{1}{2}(\Imag f)_{\Gamma\Delta}
     \nabla_\mu\left(\lambda^\Gamma\sigma^\mu\bar\lambda^\Delta\right)\right] \ ,
\end{split}
\end{equation}
with appropriate covariant derivatives $\nabla_\mu$. 
Out of the fermionic couplings we only need to recall those which allow us to
determine
the D-terms and the superpotential. They are given by
\begin{multline} \label{eq:4Dcoupl}
   \mathcal{S}_\text{Couplings}=-\frac{1}{2\kappa_4^2}\int\dd^4x\:\sqrt{-\eta} \Big[
     \sqrt{2}i\:e^{K/2}\left(\mathcal{D}_MW\chi^M\sigma^\mu\bar\psi_\mu
          +\mathcal{D}_{\bar N}\bar W\bar\chi^{\bar N}\bar\sigma^\mu\psi_\mu \right) \\
     +\text{D}_\Gamma\: \psi_\mu\sigma^\mu\bar\lambda^\Gamma 
     -\text{D}_\Gamma\: \bar\psi_\mu\bar\sigma^\mu\lambda^\Gamma+\ldots \Big] \ ,
\end{multline}
where $\ldots$ denotes all the omitted fermionic coupling terms. 

Note that all the fermionic terms \eqref{eq:4Dferm} and
\eqref{eq:4Dcoupl} of the supergravity action are determined by
supersymmetry from the bosonic action \eqref{eq:4Dbos}. On the other
hand these fermionic terms can also be used to gain insight into the
bosonic part of the supergravity action. In particular from the
couplings explicitly stated in \eqref{eq:4Dcoupl} the D-terms and
F-terms of the effective supergravity theory can be read
off. Therefore we want to compute the terms \eqref{eq:4Dcoupl} from
the super-D7-brane action. However, in order to reliably determine the
D-terms and F-terms from the D7-branes 
we first have to make sure that  the normalization of the Weyl fermions 
defined in the expansion \eqref{eq:decomptheta} are such that their 
kinetic terms agree with the kinetic terms of eq.~\eqref{eq:4Dferm}. 
This is what we will turn to next.


\subsection{Fermionic kinetic terms of the effective theory} \label{sec:fkin}


In this section we compute the fermionic kinetic terms \eqref{eq:4Dferm} of the effective four-dimensional theory for the four-dimensional gravitino and for the massless fermions arising from the super-D7-brane. First we start with the Kaluza-Klein reduced gravitino and as a second step we examine the fermionic kinetic terms resulting from the expansion of the super Dirac-Born-Infeld action.

In the string frame the Einstein-Hilbert term and the Rarita-Schwinger term appear in the ten-dimensional type IIB supergravity action as
\begin{equation}
   \mathcal{S}_\text{IIB}^\text{sf}=
      -\frac{1}{2\kappa_{10}^2}\int\dd^{10}x\sqrt{-g_{10}} e^{-2\phi} R
      -\frac{i}{2\kappa_{10}^2}\int\dd^{10}x\sqrt{-g_{10}} e^{-2\phi} 
           \pair{\bar\Psi}_M\Gamma^{MNP}\spinder_N\pair\Psi_P+\ldots \ .
\end{equation}
Performing a Kaluza-Klein reduction of the gravitinos according to \eqref{eq:4Dgrav} 
and \eqref{eq:gravitinos}, and applying a Weyl rescaling to the ten-dimensional metric
with 
\begin{align} \label{eq:weyl}
   \hat\eta=\frac{6}{\mathcal{K}}e^{\phi/2}\eta \ , && \hat g=e^{\phi/2}g \ ,
\end{align}
one obtains the four-dimensional Rarita-Schwinger term in the Einstein frame
\begin{equation} \label{eq:RS4d}
   -\frac{1}{\kappa_4^2}\int\dd^4x \sqrt{-\eta}\:\epsilon^{\mu_0\mu_1\mu_2\mu_3}\:
        \bar\psi_{\mu_0}\bar\sigma_{\mu_1}\spinder_{\mu_2}\psi_{\mu_3}
        \cdot\frac{6}{\mathcal{K}}\int_Y\dd^6y\sqrt{g}\:\check\singspin^\dagger
        \check\singspin \ .
\end{equation} 
In order to normalize this four-dimensional Rarita-Schwinger term in agreement with \eqref{eq:4Dferm} we readily read off from \eqref{eq:RS4d} with \eqref{eq:vol} the normalization of the constant internal spinors
\begin{equation} \label{eq:normsingspin}
   \check\singspin^\dagger\check\singspin=1 \ .
\end{equation}

The next task is to find the canonical normalization of the D7-brane fermions. This can be achieved by computing the fermionic kinetic terms of \eqref{eq:4Dferm}, which then sets the normalization constants in the expansion of eq.~\eqref{eq:decomptheta}. 

The kinetic terms of the massless open string sector fermions arise from the super Dirac-Born-Infeld action \eqref{eq:SDBI}. The standard four-dimensional low energy effective kinetic terms are obtained by expanding \eqref{eq:SDBI} around the fermionic fluctuations. This is achieved by computing the pullback $\super{\varphi^*}\super{g}$ of the supermetric~$\super{g}$ with respect to the supermap $\super{\varphi}$. As the starting point of this analysis we take the component expansion of the supervielbeins~$\super{E}^{\check A}_{\check M}$ of Type~IIB supergravity of ref.~\cite{Grana:2002}
\begin{equation} \label{eq:szehnb}
\begin{aligned}
   \super{E}^A_M &= E^A_M + \frac{1}{8} \bar\theta\Gamma^{ABC}\theta\:\omega_{M\:BC} \ , & \qquad
   \super{E}^\sindex{A}_M &= \Psi^\sindex{A}_M + \ldots \ , \\
   \super{E}^A_\sindex{M} &= -\frac{i}{2} \left(\bar\theta\Gamma^A\right)_\sindex{M} + \ldots \ , &
   \super{E}^\sindex{A}_\sindex{M} &= \delta_\sindex{M}^\sindex{A}+\ldots \ ,
\end{aligned}   
\end{equation}   
with the ten-dimensional spin connection $\omega_{M\:BC}$ and where $\ldots$ refers to second order terms in $\theta$ for the supervielbein components with fermionic indices. The stated order suffices for the computation of the kinetic terms. Note that we have already truncated the expressions with respect to the orientifold projection~$\mathcal{O}$. 

The next task is to compute from \eqref{eq:szehnb} the pullback supervielbein~$\super{e}^{\check A}$ given by $e^{\check A}=\dd Z^{\check M}E_{\check M}^{\check A}$ 
\begin{equation}
   \super{e}^A_M = e^A_M - \frac{i}{2}\bar\Theta\Gamma^A\spinder_M\Theta  \ , 
\end{equation}
and one obtains with $\left(\super{\varphi^*}\super{g}\right)_{MN}=\super{e}^A_M\super{e}^B_N\eta_{AB}$ the pullback supermetric to the relevant order
\begin{equation} \label{eq:smet}
   \left(\super{\varphi^*}\super{g}\right)_{MN}
     =g_{MN} - \frac{i}{2}\bar\Theta\Gamma_M\spinder_N\Theta 
             - \frac{i}{2}\bar\Theta\Gamma_N\spinder_M\Theta+\ldots \ .
\end{equation}
In addition to the fermionic fluctuations of the pullback of the ten-dimensional supermetric~$\super{g}$ we also include the bosonic fluctuations, which are computed by a normal coordinate expansion \cite{Grana:2003ek,Jockers:2004yj}
\begin{equation} \label{eq:smet2}
\begin{split}
   \super{\varphi^*}\super{g}
     =&\hat\eta_{\mu\nu}\dd x^\mu \dd x^\nu+2\hat g_{i\bar\jmath}\dd y^i\dd y^{\bar\jmath} 
       +2\hat g_{i\bar\jmath} \partial_\mu\db^i\partial_\nu\bar\db^{\bar\jmath}
        \dd x^\mu\dd x^\nu \\
      &- \frac{i}{2}\bar\Theta\Gamma_\mu\spinder_\nu\Theta \dd x^\mu\dd x^\nu
       - \frac{i}{2}\bar\Theta\Gamma_\nu\spinder_\mu\Theta\dd x^\mu\dd x^\nu+\ldots \ .
\end{split}
\end{equation}
The fields $\db^i$ and $\bar\db^{\bar\jmath}$ are the bosonic D7-brane fluctuations of the internal cycle~$S^\Lambda$, and $\db^i$ are sections of the normal bundle $H^0_+(S^\Lambda,\nbundle{S^\Lambda})$ \cite{Jockers:2004yj}, which can be expressed in terms of the basis elements~$\tilde s_A$ of Table~\ref{tab:spec} via
\begin{equation}
   \db^i=\frac{\mathcal{K}}{12i\int\Omega\wedge\bar\Omega}\:
   \bar\Omega^{ijk}\:\tilde s_{A\:jk}\db^A \ .   
\end{equation}
In the next step we insert this equation into the expansion \eqref{eq:smet2} and rescale the metric to the four-dimensional Einstein frame and obtain
\begin{equation} \label{eq:smet3}
\begin{split}
   \super{\varphi^*}\super{g}
     =&\frac{6}{\mathcal{K}}e^{\phi/2}\eta_{\mu\nu}\:\dd x^\mu \dd x^\nu
       +2 e^{\phi/2} g_{i\bar\jmath}\:\dd y^i\dd y^{\bar\jmath} \\
      &+\frac{\mathcal{K}e^{\phi/2}}{6i \int \Omega\wedge\bar\Omega}\:
       g^{j\bar l}g^{k\bar m} \:\tilde s_{A\:jk}\tilde s_{\bar B\:\bar l\bar m}\:
       \partial_\mu \db^A\partial_\nu\bar\db^{\bar B}\dd x^\mu\dd x^\nu \\
      &- \frac{i}{2}\bar\Theta\Gamma_\mu\spinder_\nu\Theta \dd x^\mu\dd x^\nu
       - \frac{i}{2}\bar\Theta\Gamma_\nu\spinder_\mu\Theta\dd x^\mu\dd x^\nu+\ldots \ .
\end{split}
\end{equation}
Furthermore the D7-brane field-strength $F$ enjoys the expansion
\begin{equation} \label{eq:F}
   F=\frac{1}{2}F_{\mu\nu}(x)\:\dd x^\mu\wedge\dd x^\nu 
     +\partial_\mu a_I(x)\:A^I\wedge\dd x^\mu
     +\partial_\mu\bar a_{\bar I}(x)\:\bar A^{\bar J}\wedge\dd x^\mu+f \ .
\end{equation}
To derive the kinetic terms of the massless D7-brane modes and the normalization of the fermionic modes we insert into the super Dirac-Born-Infeld action \eqref{eq:SDBI} the expansion \eqref{eq:smet3} and \eqref{eq:F}. Then to expand the square root in the Dirac-Born-Infeld action we use the Taylor series 
\begin{equation}
   \sqrt{\det\left(\mathfrak{A}+t\mathfrak{B}\right)}
     =\sqrt{\det\mathfrak{A}}\cdot\left[1+\frac{t}{2}\tr \mathfrak{A}^{-1}\mathfrak{B}
      +\frac{t^2}{8}\left[\left(\tr\mathfrak{A}^{-1}\mathfrak{B}\right)^2
      -2\:\tr \left(\mathfrak{A}^{-1}\mathfrak{B}\right)^2\right]+\cdots \right] \ .
\end{equation}  
Finally after inserting \eqref{eq:decomptheta} we obtain the bosonic and fermionic kinetic terms in the Einstein frame 
\begin{align} \label{eq:FermNorm}
   \mathcal{S}_\text{Kin}&=
     -\mu_7\int \dd^4 x \sqrt{-\eta}\:\int_{S^\Lambda}\dd^6\xi
      \sqrt{\det\left(e^{\phi/2}g+\dbbf^a\iota^*\omega_a - \ell\tilde f \right)} \\
     &\cdot\left[
      2\: g^{j\bar l}g^{k\bar m}\:\tilde s_{Ajk} \tilde s_{\bar B\bar l\bar m}\left(
      \frac{e^\phi}{4i\int\Omega\wedge\bar\Omega}\:
      \partial_\mu\db^A\partial^\mu\bar\db^{\bar B} 
     +i\left|N_{\chi^A}\right|^2\left(\frac{6}{\mathcal{K}}e^{-\phi/2}\right)
      \bar\chi^{\bar B}\bar\sigma^\mu\nabla_\mu\chi^A\right) \right.\nonumber \\
     &\quad+\frac{1}{4}\ell^2\:F_{\mu\nu} F^{\mu\nu} - \frac{i}{2}\left|N_\lambda\right|^2
      \left(\frac{6}{\mathcal{K}}e^{\phi/2}\right)
      \left(\lambda\sigma^\mu\nabla_\mu\bar\lambda
            +\bar\lambda\bar\sigma^\mu\nabla_\mu\lambda\right)\nonumber \\
     &\left.\quad-\frac{6}{\mathcal{K}}\bar A^{\bar I}_i A^J_{\bar\jmath}\left(
      \frac{1}{4}e^{\phi/2}\left(e^{-\phi/2}g+b-\ell f\right)^{i\bar\jmath}\:
      \partial_\mu a_I\partial^\mu\bar a_{\bar J}
     -2i\left|N_{\chi_I}\right|^2 g^{i\bar\jmath}
      \bar\chi_{\bar I}\bar\sigma^\mu\nabla_\mu\chi_J\right) \right] \ , \nonumber
\end{align}
where we have used the relation
\begin{equation} \label{eq:gid}
   \singspin^\dagger \gamma^i\gamma^j\gamma^{\bar k}\gamma^{\bar l}\singspin
     =4 \left(g^{i\bar l}g^{j\bar k}-g^{i\bar k}g^{j\bar l}\right) \ .
\end{equation}
 The normalization constants $N_\lambda$, $N_{\chi^A}$ and $N_{\chi_I}$ 
can be determined by 
comparing the 
kinetic terms  in \eqref{eq:FermNorm} with \eqref{eq:4Dbos} and 
\eqref{eq:4Dferm}. The bosonic terms determine
the K\"ahler metric and the gauge kinetic coupling function 
which then, using \eqref{eq:4Dferm}, fixes the
 $N_\lambda$, $N_{\chi^A}$ and $N_{\chi_I}$ of their superpartners.
We determine the normalization constants for $N_\lambda$ and $N_{\chi^A}$ explicitly because they are important in the following section 
\begin{align} \label{eq:norm}
   N_\lambda=\sqrt{\frac{\mathcal{K}}{6}}\ell\:e^{-\phi/4} \ , &&
   N_{\chi^A}=\frac{1}{2}\sqrt{\frac{\mathcal{K}}
     {-6i\int\Omega\wedge\bar\Omega}}\:e^{3\phi/4} \ .
\end{align}


\subsection{Fermionic D-term and F-term couplings} \label{sec:DFterms}


The couplings of the fermionic fields to the gravitino~$\psi_\mu$ reveal the structure of the D-terms and the F-terms of the theory. The former appear as couplings of the gravitino with the fermions in the vector multiplets, whereas the latter are determined from the gravitino couplings with the fermions in the chiral multiplets. The relevant terms are given in eq.~\eqref{eq:4Dcoupl}, which in our case are computed from the expansion of the super Chern-Simons action \eqref{eq:SCS} of the D7-brane.

The super RR-forms $\super{\hat C^{(q)}}$ of type IIB supergravity are stated in ref.~\cite{Bergshoeff:1999bx} where the important part for our purposes reads
\begin{equation} \label{eq:SRR}
   \super{\hat C^{(2k-2)}}=\hat C^{(2k-2)}+\frac{i}{(2k-3)!}\:e^{-\phi}\:\pair{\bar\theta}
      \mathcal{P}_k\Gamma_{M_1\ldots M_{2k-3}}\pair{\Psi}_{M_{2k-2}}
      \dd x^{M_1}\wedge\ldots\wedge\dd x^{M_{2k-2}}+\ldots \ ,
\end{equation} 
with the matrix $\mathcal{P}_k$ 
\begin{equation}
   \mathcal{P}_k = \begin{cases} \check\sigma^1 & k\ \text{even} \\
                                 i\check\sigma^2 & k\ \text{odd} \ , \end{cases}
\end{equation}
acting on the gravitino pair $\pair\psi_M$ and the coordinate pair $\pair\theta$. Note that these super RR-forms appear in ref.~\cite{Bergshoeff:1999bx} for type~II supergravity theories with O5/O9 orientifold planes. However, the orientifold compactifications needed in this paper include O3/O7 planes which are related to orientifolds with O5/O9 planes via two T-duality transformations along two distinct directions. Therefore, we also need to apply two T-duality transformations to eq.~\eqref{eq:SRR} before inserting the super RR-fields into the super Chern-Simons action \eqref{eq:SCS}. The obtained super RR-forms for O3/O7 planes are denoted by $\super{C^{(q)}}$ and \eqref{eq:SRR} becomes
\begin{equation} \label{eq:SRR2}
   \super{C^{(2k-2)}}=C^{(2k-2)}+\frac{i}{(2k-3)!}\:e^{-\phi}\:\pair{\bar\theta}
      \mathcal{P}_{k+1}\Gamma_{M_1\ldots M_{2k-3}}\pair{\Psi}_{M_{2k-2}}
      \dd x^{M_1}\wedge\ldots\wedge\dd x^{M_{2k-2}}+\ldots \ .
\end{equation} 

The couplings of \eqref{eq:4Dcoupl} arise from the super-RR-six-form field in the super Chern-Simons action \eqref{eq:SCS},
\begin{equation} \label{eq:SCS6}
   -\mu_7 \int_{\mathcal{W}} \super{\varphi^*}\left(\super{C^{(6)}}\right) \wedge \mathcal{F} \ ,
\end{equation}
with the worldvolume two-form $\mathcal{F}=B-\ell f = \dbbf - \ell\tilde f$. As before for fermionic fluctuations $\pair{\Theta}(\xi)$ the pullback of $\super{C^{(6)}}$ with respect to the supermap~$\super{\varphi^*}$ is simply obtained by replacing the odd coordinates~$\pair\theta$ in \eqref{eq:SRR2} by $\pair{\Theta}(\xi)$. Note that the pullback~$\super{\varphi^*}$ to the worldvolume~$\mathcal{W}$ also acts on the gravitino~$\pair{\psi}_M$ in eq.~\eqref{eq:SRR2}, which according to eq.~\eqref{eq:4Dgrav} then becomes 
\begin{equation} \label{eq:4Dgrav2}
\begin{split}
   \varphi^*\Psi_M\ =\ & 
     \bar\psi_\mu\otimes(\singspin\otimes\mathcal{L}^*)\otimes\mathcal{L}
     +\psi_\mu\otimes(\singspin^\dagger\otimes\mathcal{L})\otimes\mathcal{L}^* +\ldots \\
     \ =\ &\bar\psi_\mu\otimes\singspin+\psi_\mu\otimes\singspin^\dagger+\ldots \ .
\end{split}
\end{equation}
Finally taking into account the orientifold truncation \eqref{eq:gravitinos} and \eqref{eq:inv3} we obtain the pulled back super-RR-six form
\begin{equation}
   \super{\varphi^*}\left(\super{C^{(6)}}\right)=\varphi^*C^{(6)}
   +\frac{i}{5!}e^{-\phi}\:\bar\Theta\:\Gamma_{M_1\ldots M_5}
   \Psi_{M_6}\:\dd x^{M_1}\wedge\ldots\wedge\dd x^{M_6}+ \ldots \ .
\end{equation}
This equation captures couplings of the four-dimensional gravitinos $\psi_\mu$ to the D7-brane fermions $\Theta$, which are of the form \eqref{eq:4Dcoupl}, i.e. 
\begin{equation}
\begin{split}
   &\frac{i}{2\cdot 3!} e^{-\phi} \bar\Theta \Gamma_{\mu_0\mu_1\mu_2 mn}\Psi_{\mu_3}\:
   \:\dd x^{\mu_0}\wedge\ldots\wedge\dd x^{\mu_3}\wedge\dd y^m\wedge \dd y^n \\ 
  =&\frac{i}{2}\:e^{-\phi}\:\sqrt{-\hat\eta}\:\dd^4x\:\bar\Theta 
   \left(\hat\gamma^\mu\hat\gamma^5\otimes\gamma_{mn}\right)\Psi_\mu\:\dd y^m\wedge\dd y^n  \ .
\end{split}
\end{equation}
Finally inserting this expression into the super-Chern-Simons action \eqref{eq:SCS6} together with \eqref{eq:4Dgrav2} and \eqref{eq:decomptheta} we arrive after Weyl rescaling with \eqref{eq:weyl} at
\begin{align}
   \mathcal{S}_\text{Couplings}=&-\frac{i\mu_7}{2}\int\dd^4x\sqrt{-\eta}\nonumber \\
   &\cdot\left[\chi^A\sigma^\mu\bar\psi_\mu\:N_{\chi^A}\:
        \left(\tfrac{6}{\mathcal{K}}\right)^\frac{3}{2} e^{-\phi/4}
        \int_{S^\Lambda}\frac{1}{4}s_{A\:ij}\:\singspin^\dagger\gamma^j\gamma^i
        \gamma^{\bar l}\gamma^{\bar k}\singspin\:g_{l\bar l}\:g_{k\bar k}\:\dd y^l
        \wedge\dd y^k\wedge\mathcal{F} \right. \nonumber \\
   &\quad-\bar\chi^{\bar A}\bar\sigma^\mu\psi_\mu\:\bar N_{\chi^A}\:
        \left(\tfrac{6}{\mathcal{K}}\right)^\frac{3}{2} e^{-\phi/4}
        \int_{S^\Lambda}\frac{1}{4}\bar s_{\bar A\:\bar\imath\bar\jmath}\:
        \singspin^\dagger\gamma^l\gamma^k\gamma^{\bar\imath}\gamma^{\bar\jmath}\singspin\:
        g_{l\bar l}\:g_{k\bar k}\:\dd y^{\bar l}\wedge\dd y^{\bar k}
        \wedge\mathcal{F}\nonumber \\
   &\quad+\psi_\mu\sigma^\mu\bar\lambda\:\bar N_\lambda\:
        \left(\tfrac{6}{\mathcal{K}}\right)^\frac{3}{2} e^{\phi/4}
        \int_{S^\Lambda}P_-\:\singspin^\dagger\gamma^i\gamma^{\bar\jmath}\singspin\:
        g_{i\bar k}g_{l\bar\jmath}\:\dd y^{\bar k}\wedge\dd y^l
        \wedge\mathcal{F}\nonumber \\
   &\quad-\left.\bar\psi_\mu\bar\sigma^\mu\lambda\: N_\lambda\:
        \left(\tfrac{6}{\mathcal{K}}\right)^\frac{3}{2} e^{\phi/4}
        \int_{S^\Lambda}P_-\:\singspin^\dagger\gamma^i\gamma^{\bar\jmath}\singspin\:
        g_{i\bar k}g_{l\bar\jmath}\:\dd y^{\bar k}\wedge\dd y^l\wedge\mathcal{F} \right] \ .
\end{align}
This can be rewritten by applying the Dirac gamma matrix identities $\ac{\gamma^i}{\gamma^{\bar\jmath}}=2g^{i\bar\jmath}$ and \eqref{eq:gid}. Furthermore we insert the normalization constants \eqref{eq:norm} and the definition of the K\"ahler potential~\eqref{eq:K1} to cast these couplings into the form
\begin{align} \label{eq:CouplCS}
   \mathcal{S}_\text{Couplings}=&-\frac{1}{2\kappa_4^2}\int\dd^4x\sqrt{-\eta} \nonumber \\
   &\cdot\left[\sqrt{2}i\:e^{K/2}\cdot \kappa_4^2\mu_7\left(
        \chi^A\sigma^\mu\bar\psi_\mu\:\int_{S^\Lambda}\tilde s_A\wedge\mathcal{F}
        +\bar\chi^{\bar A}\bar\sigma^\mu\psi_\mu 
        \int_{S^\Lambda}\tilde s_{\bar A}\wedge\mathcal{F}\right)\right. \nonumber \\
   &\quad+\left.\left(\psi_\mu\sigma^\mu\bar\lambda-\bar\psi_\mu\bar\sigma^\mu\lambda\right)
        \left(-12\kappa_4^2\mu_7\ell\frac{1}{\mathcal{K}}\int_{S^P}J\wedge\mathcal{F}\right)
        \right] \ .
\end{align}
Now comparing these fermionic couplings with eq.~\eqref{eq:4Dcoupl}
one extracts immediately a generic expression for 
the D-term of the $U(1)$ gauge theory originating from the D7-brane
\begin{equation}\label{eq:Dfermionic}
   \text{D}\ =\ \frac{12\kappa_4^2\mu_7\ell}{\mathcal{K}}\int_{S^P}J\wedge\mathcal{F} \
    =\
\frac{12\kappa_4^2\mu_7\ell}{\mathcal{K}}
    \left(\mathcal{K}_{Pa}\dbbf^a-Q_\alpha v^\alpha\right) \ ,
\end{equation}
where in the last step we used again \eqref{eq:K} and \eqref{eq:notvanish}.
Hence the fermionic D7-brane reduction confirms the D-term
 \eqref{eq:fDterm}, which is computed in section~\ref{sec:fluxes} by
 means of analyzing the bosonic part of the 
 $\mathcal{N}=1$ supergravity action. Note that the fermionic
 computation has neither required the knowledge of the gauged
 isometries
 nor the structure of the K\"ahler potential. However, for the
 supergravity derivation it is crucial to know the definition of the
 chiral variables and the K\"ahler potential, as this information
 enters in the differential equation \eqref{eq:Dterms} which encodes
 the D-term. Therefore the fermionic computation is an
alternative and more direct way to determine the D-term.
For the case at hand it confirms the supergravity computation
and in addition 
checks the definition of the K\"ahler potential \eqref{eq:K1}. 

In a second step we match the terms in \eqref{eq:4Dcoupl} with \eqref{eq:CouplCS} and we readily determine the flux induced superpotential. The integral relation \eqref{eq:vanishint} is responsible for the superpotential to be independent of $\dbbf$ because the two-forms $\tilde s_A$ are elements of $\tilde H^2_-(S^\Lambda)$ whereas the two-form $\dbbf$ are inherited from the bulk. Therefore we arrive at the holomorphic superpotential
\begin{equation} \label{eq:W1}
   W(\db)\ =\ \kappa_4^2\mu_7\, Q_A \db^A \ , \qquad
Q_A = \ell \int_{S^\Lambda} \tilde s_A \wedge \tilde f \ .
\end{equation} 
This superpotential is determined from \eqref{eq:CouplCS} by reading off and integrating $\partial_A W$. However, the fermionic couplings \eqref{eq:4Dcoupl} tell us that the low energy effective action should not only contain the term proportional to $\partial_A W$ but also the term proportional to $K_A\:W$, which then completes $\partial_A W$ to the K\"ahler covariant derivative $\mathcal{D}_A W$. The couplings proportional to $K_A\:W$ written in the ten-dimensional string frame become 
\begin{multline} \label{eq:covterm}
   \frac{1}{2\kappa_4^2}\int\dd^4x\:\sqrt{-\eta}
    \sqrt{2}i\:e^{K/2}K_MW\chi^M\sigma^\mu\bar\psi_\mu \\
    \xrightarrow{\text{\ string frame \ }} 
   \ \mu_7\ell \int\dd^4x\sqrt{-\hat\eta}\: \chi^A\sigma^\mu\bar\psi_\mu\:
    \sqrt{\frac{\mathcal{K}}{-6i\int\Omega\wedge\bar\Omega}}\:\mathcal{L}_{A\bar B}
    \bar\db^{\bar B}\:W \ . 
\end{multline}
This expression brings about two important observations: First of all it involves two integrals over the internal D7-brane cycle~$S^\Lambda$, namely one integral is hidden in the definition \eqref{eq:LC} of $\mathcal{L}_{A\bar B}$ and the other integral appears in the superpotential~$W$ as stated in \eqref{eq:W1}. However, these two integrals can never be generated from the Chern-Simons action~\eqref{eq:SCS}. Second, the Dirac-Born-Infeld action~\eqref{eq:SDBI}, which captures only open string tree-level amplitudes, is weighted with the dilaton factor $e^{-\phi}$. This is due to the fact that open string tree-level amplitudes have Euler characteristic one. The term \eqref{eq:covterm}, however, does not contain (in the string frame) a factor of dilaton. Therefore it is not obtained by the normal coordinate expansion of the open string tree-level Dirac-Born-Infeld action. In principal such a term can be generated at the open string one loop level because the cylinder amplitude and the M\"obius amplitude with Euler characteristic zero is not weighted by a factor of dilaton.

Before we conclude this section, we want to come back to the computed superpotential \eqref{eq:W1}. In order to gain further insight we need to review some aspects of the complex structure moduli space in the presence of D7-branes as studied in detail in ref.~\cite{Jockers:2004yj}. So far in the analysis we have treated the D7-brane fluctuations~$\db$ and the complex structure deformations~$z$ independently. However, the complex structure of the D7-brane cycle~$S^\Lambda$ is inherited from the ambient space~$Y$. Therefore only in the limit of small brane fluctuations~$\db$ and small bulk deformations~$z$ these fields are not interlinked. In order to generalize the results away from this limit it is necessary to treat the fields~$\db$ and $z$ in the common moduli space~$\mathcal{M}_{\mathcal{N}=1}$, which is the moduli space resulting from the variation of Hodge structure of the relative cohomology group $H^3_-(Y,S^\Lambda)$ \cite{Mayr:2001,Jockers:2004yj}. This cohomology group can be identified with
\begin{equation} \label{eq:RelCoh}
   H_-^3(Y,S^\Lambda)\cong\tilde H^3_-(Y)\oplus\tilde H^2_-(S^\Lambda) \ ,
\end{equation}
where $\tilde H^3_-(Y)=\ker (H^3_-(Y)\xrightarrow{\iota^*} H^3_-(S^\Lambda))$ and where $\tilde H^2_-(S^\Lambda)$ is defined in eq.~\eqref{eq:FluxCoh}. Now the complex structure deformations~$z$ are identified with elements of $\tilde H^3_-(Y)$ whereas the D7-brane fluctuations~$\db$ expanded into two-forms $\tilde s_A$ correspond to elements of $\tilde H^2_-(S^\Lambda)$. Both kind of fields combine into the relative cohomology group $H^3(Y,S^\Lambda)$. We underline elements of $H^3_-(Y,S^\Lambda)$, e.g. $\rel{\Theta}$, and introduce the projection operators $P^{(3)}$ and $P^{(2)}$ onto the three-form part and two-form part of \eqref{eq:RelCoh} respectively. Now in this generalized complex structure moduli space~$\mathcal{M}_{\mathcal{N}=1}$ the D7-brane fluctuations are expanded into relative two-forms~$\rel{\tilde s_A}$ and the superpotential \eqref{eq:W1} becomes
\begin{equation} \label{eq:W2}
   W(\db,z)=\kappa_4^2\mu_7\, Q_A(z)\, \db^A \ , \qquad
Q_A(z) = \ell \int_{S^\Lambda} P^{(2)}\rel{\tilde s_A}\: 
    \wedge \tilde f \ .
\end{equation}
Note that the superpotential~\eqref{eq:W2} in the moduli space~$\mathcal{M}_{\mathcal{N}=1}$ couples not only to the D7-brane fluctuations~$\db$ but also to the bulk complex structure moduli~$z$ because the variation of Hodge structure for relative forms tells us that
\begin{equation} \label{eq:Kod2}
   \partial_{z^{\tilde a}} \rel{\tilde s_A} = k_{\tilde a} \rel{\tilde s_A} + \rel{\eta_{\tilde aA}} \ .
\end{equation}
where $\left.P^{(2)}\rel{\eta_{\tilde aA}}\right|_{z,\db=0}\in\tilde H^{(1,1)}_{\bar\partial,-}(S^\Lambda)$. Finally let us come back to the F-term fluxes $\tilde f$. In terms of the complex structure for a generic cycle~$S^\Lambda$ these fluxes split into a $(2,0)$ and $(0,2)$ part and into a $(1,1)$ part.\footnote{The $(1,1)$-part in $\tilde H^2_-(S^\Lambda)$ should not be confused with the $(1,1)$-fluxes~$\leftup{Y}{f}$ of $\leftup{Y}{H^2_-(S^\Lambda)}$.} The $(0,2)$ fluxes couple already in the superpotential~\eqref{eq:W1} to the D7-brane fluctuations~$\db$ because the $\tilde s_A$ are $(2,0)$-forms. However, $(1,1)$ fluxes do not enter in the expression \eqref{eq:W1} but only in the $\mathcal{M}_{\mathcal{N}=1}$ superpotential \eqref{eq:W2} where these fluxes couple in first order to the complex structure moduli~$z$.

Thus we see that in general the fluxes $\tilde f$ contribute to the D-term \eqref{eq:Dfermionic} and in addition enter the superpotential \eqref{eq:W2}. This property of D7-brane fluxes has already been observed in ref.~\cite{Lust:2005bd} in the context of F-theory compactifications on $K3\times K3$, where it is shown that background fluxes correspond to both D- and F-terms in the scalar potential after truncating to an effective $\mathcal{N}=1$ theory. In the corresponding orientifold limit to $T^2/\mathbf{Z}_2\times K3$ with D7-branes \cite{Sen:1996vd} some of these fluxes become D7-brane fluxes which exhibit also this property \cite{Lust:2005bd}. 


\subsection{Superpotential from holomorphic Chern-Simons action} \label{sec:HCS}


In this section we want to describe how the superpotential~\eqref{eq:W2} can also be derived from the topological B-model \cite{Witten:1991zz}. Topological string theories can be obtained by twisting the two-dimensional $\mathcal{N}=(2,2)$ non-linear sigma model of the worldsheet. This twist transforms two linear combination of sigma model supercharges into BRST operators of the resulting topological theory. There are two inequivalent twists namely the A- and B-twist. In our case the B-model is relevant because it allows us to describe D-branes wrapping holomorphic cycles. This is due to the fact that D-branes wrapping holomorphic cycles preserve the linear combination of supercharges which become BRST operators in the B-model \cite{Hori:2000ck}.

In ref.~\cite{Witten:1992fb} it is shown that the topological open string disk partition function in the presence of $N$ D-branes wrapping the entire internal Calabi-Yau manifold~$Y$ is given by the holomorphic Chern-Simons action 
\begin{equation} \label{eq:HCS}
   W_Y=\int_Y\Omega\wedge\tr \left(A\wedge\bar\partial A +\frac{2}{3}A\wedge A\wedge A\right) \ , 
\end{equation}
where $A$ is the gauge field of the $U(N)$ gauge theory of the D-branes. On the other hand the open string disk partition function is the superpotential in the low energy effective action of the physical string theory \cite{Bershadsky:1993cx}. Thus in order to obtain the superpotential for the D7-brane the holomorphic Chern-Simons action~\eqref{eq:HCS} for the six-cycle~$Y$ must be dimensionally reduced to the holomorphic four cycle~$S^\Lambda$. This is achieved by saturating the normal components of the integrand \eqref{eq:HCS} with the D7-brane fluctuations~$\db^i$ which are sections in the normal bundle of $S^\Lambda$ \cite{Kachru:2000ih,Mayr:2001,Lust:2005bd}, i.e 
\begin{equation} \label{eq:W3}
   W_{S^\Lambda}=\frac{1}{2}\int_{S^\Lambda}\Omega_{ijk}\db^i\:
     \partial_{\bar\imath} A_{\bar\jmath}\:
     \dd z^j\wedge\dd z^k\wedge\dd\bar z^{\bar\imath}\wedge\dd\bar z^{\bar\jmath} \ .
\end{equation} 
Note that for a single D7-brane the non-Abelian second term of \eqref{eq:HCS} vanishes. The obtained superpotential~\eqref{eq:W3} can be further simplified because $\tfrac{1}{2}\Omega_{ijk}\db^i\dd z^j\wedge\dd z^k=\tilde s_A\db^A$ \cite{Jockers:2004yj}, and $\partial_{\bar\imath}A_{\bar\jmath}\dd\bar z^{\bar\imath}\wedge\dd\bar z^{\bar\jmath}$ is just the $(0,2)$ part of the background flux~$\tilde f$. Therefore eq.~\eqref{eq:W3} simplifies with \eqref{eq:vanishint} to
\begin{equation}
   W_{S^\Lambda}\ \sim\ Q_A \db^A\ ,
\end{equation} 
in agreement with \eqref{eq:W1}. 

Finally let us consider a (small) complex structure deformation~$z^{\tilde a}$, which modifies the definition of the holomorphic three-form $\Omega$, that is to say the deformed holomorphic three-form differs to lowest order from the undeformed three-from by the $(2,1)$-form $iz^{\tilde a}\chi_{\tilde a}$, which can be seen from the Kodaira formula $\partial_{z^{\tilde a}}\Omega=k_{\tilde a}\Omega + i \chi_{\tilde a}$ \cite{Candelas:1990pi}. Thus also including small complex structure deformations~$z$ the reduction of the holomorphic Chern-Simons action \eqref{eq:HCS} leads to 
\begin{equation} \label{eq:W4}
   W_{S^\Lambda}=\frac{1}{2}\int_{S^\Lambda}\left[\left(1+z^{\tilde a}k_{\tilde a}\right)
     \Omega_{ijk}\db^i\:\partial_{\bar\imath} A_{\bar\jmath}
     +iz^{\tilde a}\chi_{\tilde a\:ij\bar\imath}\db^i\:\partial_{\bar\jmath}A_k\right]
     \dd z^j\wedge\dd z^k\wedge\dd\bar z^{\bar\imath}\wedge\dd\bar z^{\bar\jmath} \ .
\end{equation} 
The contractions of the three-forms $\Omega$ and $\chi_{\tilde a}$ with the holomorphic normal bundle section~$\db^i$ generates a two-form in $\tilde H^2_-(S^\Lambda)$.\footnote{The resulting two-form is not inherited from the Calabi-Yau~$Y$ because in the deformed complex structure it corresponds to a $(2,0)$-form which is an element of $H^2_-(S^\Lambda)$.} Furthermore comparing for small deformations~$z^{\tilde a}$ with \eqref{eq:Kod2} we identify the resulting two-form with the projected relative two-form $P^{(2)}\left((1+k_{\tilde a}z^{\tilde a})\rel{\tilde s_A}+z^{\tilde a}\rel{\eta_{\tilde a A}}\right)\db^A$. Therefore we arrive for the superpotential ~\eqref{eq:W4} at
\begin{equation}
   W_{S^\Lambda}\ \sim\ Q_A(z)\,\db^A \ ,
\end{equation}
where we have again applied the integral relation~\eqref{eq:vanishint} because $P^{(2)}\rel{\tilde s_A}$ is an element of $\tilde H^2_-(S^\Lambda)$ and therefore there are no couplings to $\leftup{Y}{f}$. Note that this superpotential obtained from the reduction of the holomorphic Chern-Simons action is in agreement with the previous computed superpotential in eq.~\eqref{eq:W2}.


\section{Explicit examples} \label{sec:scalarpot}


In order to gain more insight into the 
scalar potential let us analyze two instructive examples by computing  their scalar potentials explicitly. In section~\ref{sec:onet} we discuss a model which has just one K\"ahler modulus, whereas in section~\ref{sec:threet} we choose a setup which resembles a toroidal compactification or some orientifolded version thereof.


\subsection{One K\"ahler modulus} \label{sec:onet}


Our first example features a Calabi-Yau orientifold~$Y$ with
$h_+^{1,1}=1$ and with $h_-^{1,1}=0$, that is to say we have a single
harmonic two-form $\omega_\Lambda$. In order to obtain convenient
numerical factors we choose the single triple intersection number to
be $\mathcal{K}_{\Lambda\Lambda\Lambda}=6$. The K\"ahler variables in
the theory are given by $S$, $\db^A$, and $T^\Lambda$, that is we also
keep the complex structure deformations fixed. We turn on D7-brane
background fluxes such that $Q_\Lambda$ and $\tilde f$ are non-zero
but $Q_{\tilde f}$ vanishes. For this particular model the K\"ahler
potential \eqref{eq:K1} can be stated explicitly in terms of its
K\"ahler variables and becomes\footnote{We do not have an explicit
  Calabi-Yau orientifold with all these required properties. However,
  this example is instructive because it reveals many features of the
  generic case discussed in the previous chapters.}
\begin{equation} \label{eq:Konet}
   K(S,T_\Lambda,\db^A)=
     -\log\left[-i(\dil-\bar\dil)+2i\kappa_4^2\mu_7\mathcal{L}_{A\bar B}\db^A\bar\db^B\right]
     -3\log\tfrac{1}{9}\left[T_\Lambda+\bar T_\Lambda\right] \ .
\end{equation}
Due to the non-vanishing charge $Q_\Lambda$ the K\"ahler modulus $T_\Lambda$ becomes charged under the $U(1)$ of the D7-brane and transforms non-linearly according to \eqref{eq:covT}. The corresponding D-term scalar potential of eq.~\eqref{eq:Vf} reduces for this particular setup to
\begin{equation} \label{eq:spotD}
   V_\text{D}=\frac{6\kappa_4^2\mu_7}{T_\Lambda+\bar T_\Lambda}
      \left(\frac{9\:Q_\Lambda}{T_\Lambda+\bar T_\Lambda}-\sum_i 
      q_i \abs{X_i}^2\right)^2 \ ,
\end{equation}   
where using \eqref{eq:DtermShift} we have included the additional
charged chiral matter multiplets $X_i$ with charge $q_i$ arising from
D7-brane intersections as discussed in section~\ref{sec:fluxes}. 
This form of the scalar potential precisely coincides 
with the potential obtained in ref.~\cite{Burgess:2003ic}. 
However, we should stress that it crucially depends 
on the existence of a  non-vanishing $Q_\Lambda$ and the absence
of $\dbbf$ which follows from our choice $h_-^{1,1}=0$.

As already discussed in ref.~\cite{Burgess:2003ic}
the  minimum of  $V_\text{D}$ depends on the properties
of other couplings of $X_i$ and also on possible non-perturbative corrections.
If the vacuum expectation value of the $X_i$ is not fixed by
additional F-term couplings , $V_\text{D}=0$ can be obtained by 
adjusting $\langle X_i\rangle$. If, on the other hand, F-terms impose
$\langle X_i\rangle=0$ a vanishing D-term potential 
only occurs for $T_\Lambda+\bar T_\Lambda \rightarrow \infty$
resulting in a run-away behavior. However, as discussed in 
\cite{KKLT:2003,Gorlich:2004qm,Denef:2004dm} 
the K\"ahler modulus $T_\Lambda$ can be stabilized by non-perturbative
effects such as Euclidean-D3-brane instantons and/or gaugino
condensation on a stack of D7-branes. In this case the D-term
spontaneously breaks supersymmetry and can indeed
 provides for a mechanism to uplift an Anti-deSitter vacuum to a
 metastable deSitter minimum along the lines of ref.\ \cite{KKLT:2003}.

Now we turn to the discussion of the F-term scalar potential, which is computed by inserting \eqref{eq:Konet} and \eqref{eq:W1} into \eqref{eq:spot}, i.e.
\begin{equation}
   V_\text{F}=\tfrac{3^6}{2} \kappa_4^2\mu_7\, G^{C\bar D} Q_C Q_{\bar D} \,
     \frac{1+\kappa_4^2\mu_7 e^{\phi} G_{A\bar B}\:\db^A\bar\db^B}
     {(T_\Lambda+\bar T_\Lambda)^3}\, \ .
\end{equation}
Here we have defined the metric $G_{A\bar B}=i\mathcal{L}_{A\bar B}$
and its inverse $G^{A\bar B}$. As one can easily see the effect of the
F-term scalar potential is twofold. On the one hand it also exhibits a
runaway behavior but on the other hand once the K\"ahler modulus~$T_\Lambda$ is stabilized the fluxes~$\tilde f$ render some of the fluctuations~$\db^A$ massive and hence stabilize these D7-brane fields.


\subsection{Three K\"ahler moduli} \label{sec:threet}


In a second example we consider a Calabi-Yau orientifold with the
Hodge numbers $h_+^{1,1}=3$ and $h_-^{1,1}=0$, namely with three
positive harmonic two-forms $\omega_\Lambda$, $\omega_1$ and
$\omega_2$. Moreover up to permutations the only non-vanishing
intersection number is $\mathcal{K}_{\Lambda 1 2}=1$. The D7-brane
field content is given by two Wilson line multiplets $a_1$ and $a_2$
and one D7-brane matter multiplet $\db$. Furthermore the non-zero
D7-brane couplings are chosen to be $i\mathcal{C}^{1\bar
  1}_1=i\mathcal{C}^{2\bar 2}_2=i\mathcal{L}=1$. Note that this field
content and the specified intersection numbers resembles the structure
of a (orientifolded) six-torus as discussed in
refs.~\cite{Lust:2004cx,Lust:2004}. The non-vanishing flux charges are
$Q_1$, $Q_2$ and $\tilde f$ but we keep 
$Q_{\tilde f}=0$.  Then the K\"ahler potential \eqref{eq:K1} in terms of the K\"ahler variables $S$, $T_\Lambda$, $T_1$, $T_2$, $a_1$, $a_2$ and $\db$ becomes
\begin{align}
   K(S, T, a, \db)=&-\log\left[-i(S-\bar S)-2\kappa_4^2\mu_7\db\bar\db\right] 
     -\log\tfrac{2}{3}\left[T_\Lambda+\bar T_\Lambda\right]  \\
     &-\log\tfrac{2}{3}\left[T_1+\bar T_1-6\kappa_4^2\mu_7\ell^2 a_1\bar a_1\right]
     -\log\tfrac{2}{3}\left[T_2+\bar T_2-6\kappa_4^2\mu_7\ell^2 a_2\bar a_2\right] \nonumber \ .
\end{align}
Then analogously to the computation in section~\ref{sec:onet} the scalar potential for this setup becomes
\begin{equation}
   V_\text{D}=\frac{27\kappa_4^2\mu_7}{2(T_\Lambda+\bar T_\Lambda)}
     \left(\frac{Q_1}{T_2+\bar T_2-6\kappa_4^2\mu_7\ell^2 a_2\bar a_2}
     +\frac{Q_2}{T_1+\bar T_1-6\kappa_4^2\mu_7\ell^2 a_1\bar a_1}\right)^2 \ ,
\end{equation}
whereas the F-term scalar potential computed with \eqref{eq:spot} and \eqref{eq:W1} reads
\begin{equation}
   V_\text{F}=
      \frac{3^3\kappa_4^2\mu_7(1+e^\phi\kappa_4^2\mu_7\db\bar\db)}
           {2^3(T_\Lambda+\bar T_\Lambda)(T_1+\bar T_1-6\kappa_4^2\mu_7\ell^2 a_1\bar a_1)
            (T_2+\bar T_2-6\kappa_4^2\mu_7\ell^2 a_2\bar a_2)} 
      \left(\ell\int_{S^\Lambda}\tilde s\wedge\tilde f\right)^2 \ .
\end{equation}
Note that qualitatively this example exhibits similar features as the case studied in the previous section, namely the D- and F-term potentials drive the theory to some decompactification limit and in addition the F-term renders the D7-brane field~$\zeta$ massive.


\section{Conclusions} \label{sec:conc}


In this paper we discussed a space-time filling D7-brane in Calabi-Yau
orientifolds with non-trivial background fluxes for the $U(1)$ gauge
theory localized on the D7-brane worldvolume.  We found that these
fluxes induce D- and F-terms in the four-dimensional effective
$\mathcal{N}=1$ supergravity description. As it is difficult to obtain
the scalar potential terms from the Kaluza-Klein reduction of the
bosonic Dirac-Born-Infeld and Chern-Simons action of the D7-brane we
concentrated on particular fermionic terms which allowed
us to reliably determine the D- and F-terms. We found that both can be expressed
generically in terms
of worldvolume integrals containing the background fluxes.
Furthermore the computed D-term is in agreement 
with the supergravity analysis which uses the Killing vector of the
gauged isometry to infer
the D-term. We also showed that the F-term can be obtained by 
a dimensional reduction of the holomorphic Chern-Simons action. 

The D7-brane fluxes naturally split into a contribution
$\leftup{Y}f$
which can be expanded into two-forms inherited from the ambient
Calabi-Yau space and into two-forms $\tilde f$
which are intrinsic harmonic forms on the wrapped D7-brane cycle. The 
fluxes $\leftup{Y}f$ have the effect of 
redefining some chiral variables and thereby adjusting the D-term,
which is already present without turning on D7-brane fluxes. In the
minimized scalar potential this modification amounts to a shift 
in the vacuum expectation values of the $b^a$ fields. The effect of 
$\tilde f$ is more divers in that they can contribute to both F- and
D-terms. Whenever the worldvolume integrals 
$Q_A$ defined in \eqref{eq:W1} are non-zero a linear
superpotential for the  D7-brane matter fields arises generating 
mass-like terms.
The fluxes $\tilde f$ also 
contribute to the D-term whenever the worldvolume integrals $Q_\alpha$
defined in \eqref{eq:notvanish} do not vanish. In this case 
the K\"ahler moduli $T_\alpha$ become charged under the $U(1)$ gauge
symmetry.
A special role is played by a non-vanishing $Q_\Lambda$ defined in
\eqref{QLdef} when extra charged chiral matter fields appear
at the intersection of the D7-brane with its orientifold image.
This renders the $U(1)$ anomalous and results in a further correction
to the D-term. However,  non-vanishing $Q_\Lambda$ appears to be a
geometrical condition which is not easily satisfied.

In ref.~\cite{Burgess:2003ic} D7-brane background fluxes have been
suggested as a source for a  positive cosmological constant. 
Our analysis confirms this proposal but
only for a very specific type of D7-brane fluxes and a very special
class of Calabi-Yau orientifolds. However, a more detailed analysis
is necessary to show the existence of a metastable minimum.
Certainly the K\"ahler potentials of these simple models are further corrected by taking into account the combined complex structure and D7-brane moduli space \cite{Mayr:2001,Jockers:2004yj} and by including the back-reaction of D7-branes to geometry. These aspects were beyond the scope of this work and they deserve further study. Maybe these questions are best addressed by studying F-theory compactifications on Calabi-Yau fourfolds along the lines of refs.~\cite{Haack:2001jz,Lust:2005bd}.


\vskip 1cm
\bigskip
\noindent {\Large{\bf Appendix}}
\appendix


\section{$\mathcal{N}=1$ low energy effective action} \label{sec:action}



\subsection{Effective action without D7-brane fluxes} \label{sec:noaction}


The bosonic part of the low energy effective action resulting from type~IIB string theory compactified on the Calabi-Yau orientifold~$Y$ in the presence of a space-time filling D7-brane is obtained form a Kaluza-Klein reduction. The starting point of the reduction is the ten-dimensional type~IIB supergravity action for the bulk fields, whereas the D7-brane fields are deduced from the eight-dimensional Dirac-Born-Infeld and Chern-Simons worldvolume action. The Dirac-Born-Infeld action of a single D7-brane is given in the string frame by
\begin{align} \label{eq:DBI}
   \mathcal{S}^{\text{sf}}_{\text{DBI}}
      =-\mu_7\int_\mathcal{W}\dd^7\xi
       \:e^{-\phi}\sqrt{-\det \left(\varphi^*(g_{10}+B)_{ab}-\ell F_{ab}\right)} \ , 
   && \ell=2\pi\alpha' \ ,
\end{align}
and its Chern-Simons action reads
\begin{equation} \label{eq:CS}
   \mathcal{S}_{\text{CS}}=\mu_7 \int_{\mathcal{W}}
       \sum_q \varphi^*\left(C^{(q)}\right)e^{\ell F-\varphi^*B} \ .
\end{equation}
Here the constant $\mu_7$ is the tension of the D7-brane and due to the BPS-condition it is also its RR-charge.
 
The Kaluza-Klein reduction to four-dimensions is presented in detail in ref.~\cite{Jockers:2004yj} and the result of this analysis reads
\begin{align} \label{eq:action}
   \mathcal{S}^{\text{E}}
     =&\frac{1}{2\kappa_4^2}\int \left[-R\:*_41
          +2\mathcal{G}_{\tilde a\tilde b}\dd z^{\tilde a} \wedge *_4 \dd\bar z^{\tilde b}
          +2G_{\alpha\beta}\dd v^\alpha \wedge *_4 \dd v^\beta \right. \nonumber \\
      &+\frac{1}{2}\dd(\ln \mathcal{K})\wedge *_4 \dd(\ln \mathcal{K})
          +\frac{1}{2}\dd\phi\wedge *_4 \dd\phi 
          +2e^\phi G_{ab}\dd b^a\wedge *_4 \dd b^b \nonumber \\
      &+2i\kappa_4^2\mu_7\mathcal{L}_{A\bar B}\left(e^\phi+4 G_{ab} b^a b^b\right)
          \dd\db^A\wedge *_4\dd\bar\db^{\bar B} 
          +\frac{24}{\mathcal{K}}\kappa_4^2\mu_7\ell^2i\mathcal{C}^{I\bar J}_\alpha v^\alpha
          \dd a_I\wedge *_4\dd\bar a_{\bar J} \nonumber \\
      &+\frac{e^{2\phi}}{2}
          \left(\dd l+\kappa_4^2\mu_7\mathcal{L}_{A\bar B}
          \left(\dd\db^A\bar\db^{\bar B}-\dd\bar\db^{\bar B}\db^A\right)\right)\wedge  
          *_4\left(\dd l+\kappa_4^2\mu_7\mathcal{L}_{A\bar B}
          \left(\dd\db^A\bar\db^{\bar B}-\dd\bar\db^{\bar B}\db^A\right)\right) \nonumber \\
      &+2e^\phi G_{ab}
          \left(\cov c^a-l\dd b^a-\kappa_4^2\mu_7 b^a \mathcal{L}_{A\bar B}
          \left(\dd\db^A\bar\db^{\bar B}-\dd\bar\db^{\bar B}\db^A\right)\right)\wedge \nonumber \\
      &\qquad *_4 \left(\cov c^b-l\dd b^b-\kappa_4^2\mu_7 b^b 
          \mathcal{L}_{A\bar B} 
          \left(\dd\db^A\bar\db^{\bar B}-\dd\bar\db^{\bar B}\db^A\right)\right) \nonumber \\
      &+\frac{9}{2\mathcal{K}^2}G^{\alpha\beta}
          \left(\cov\rho_\alpha-\mathcal{K}_{\alpha bc}
          c^b\dd b^c-\tfrac{1}{2}\kappa_4^2\mu_7
          \mathcal{K}_{\alpha bc} b^b b^c\mathcal{L}_{A\bar B}
          \left(\dd\db^A\bar\db^{\bar B}-\dd\bar\db^{\bar B}\db^A\right)\right. \nonumber \\
      &\qquad\qquad\left. +2\kappa_4^2\mu_7\ell^2\mathcal{C}_\alpha^{I\bar J}
          \left(a_I\dd \bar a_{\bar J}-\bar a_{\bar J}\dd a_I\right)\right)\wedge \nonumber \\
      &\qquad *_4 \left(\cov\rho_\beta-\mathcal{K}_{\beta ab} c^a\dd b^b
          -\tfrac{1}{2}\kappa_4^2\mu_7
          \mathcal{K}_{\beta bc}b^b b^c\mathcal{L}_{A\bar B}
          \left(\dd\db^A\bar\db^{\bar B}-\dd\bar\db^{\bar B}\db^A\right)\right. \nonumber \\
      &\qquad\qquad\left. +2\kappa_4^2\mu_7\ell^2\mathcal{C}_\beta^{I\bar J}
          \left(a_I\dd\bar a_{\bar J}-\bar a_{\bar J}\dd a_I\right)\right) \nonumber \\
      &+\kappa_4^2\mu_7\ell^2 \left(\tfrac{1}{2}\mathcal{K}_\Lambda
          -\tfrac{1}{2}e^{-\phi}\mathcal{K}_{\Lambda ab} b^a b^b 
          \right) F\wedge *_4F \nonumber \\
      &+\kappa_4^2\mu_7\ell^2\left(\rho_\Lambda-\mathcal{K}_{\Lambda ab}c^a b^b
          +\tfrac{1}{2}\mathcal{K}_{\Lambda ab}b^a b^bl \right) F\wedge F \nonumber \\
      &\left.+\frac{1}{2}(\Imag \mathcal{M})_{\hat\alpha\hat\beta} 
          \dd V^{\hat\alpha}\wedge *_4 \dd V^{\hat\beta}
          +\frac{1}{2}(\Real \mathcal{M})_{\hat\alpha\hat\beta}
          \dd V^{\hat\alpha}\wedge\dd V^{\hat\beta} \right] \ . 
\end{align}
$\kappa_4$ is now the four-dimensional gravitational coupling constant and
\begin{align} \label{eq:triple}
   \mathcal{K}_{\alpha\beta\gamma}
     =\int_Y \omega_{\alpha}\wedge\omega_{\beta}\wedge\omega_{\gamma} \ , 
   &&\mathcal{K}_{ab\gamma}
     =\int_Y \omega_a\wedge\omega_b\wedge\omega_\gamma \ ,
\end{align}
are the non-vanishing triple intersection numbers of the Calabi-Yau manifold~$Y$. Additionally we abbreviate contractions of these intersection numbers with the fields $v^\alpha$ and obtain with \eqref{eq:NS} the non-vanishing combinations 
\begin{equation} \label{eq:K}
\begin{aligned}
   \mathcal{K}&=\int_Y J\wedge J\wedge J
      =\mathcal{K}_{\alpha\beta\gamma}v^\alpha v^\beta v^\gamma \ , 
   & \mathcal{K}_\alpha&=\int_Y \omega_\alpha\wedge J\wedge J
      =\mathcal{K}_{\alpha\beta\gamma}v^\beta v^\gamma \ , \\
   \mathcal{K}_{\alpha\beta}&=\int_Y \omega_\alpha\wedge\omega_\beta\wedge J
      =\mathcal{K}_{\alpha\beta\gamma}v^\gamma \ , 
   & \mathcal{K}_{ab}&=\int_Y \omega_a\wedge\omega_b\wedge J
      =\mathcal{K}_{ab\gamma}v^\gamma \ .
\end{aligned}
\end{equation}
Note that $\mathcal{K}$ is proportional to the volume of the internal Calabi-Yau manifold~$Y$, i.e.
\begin{equation} \label{eq:vol}
    \vol (Y)=\frac{\mathcal{K}}{6}  \ .
\end{equation}

In the action \eqref{eq:action} there appear also various metrics. On the space of harmonic two-forms one defines the metrics \cite{Strominger:1985ks,Candelas:1990pi}
\begin{equation} \label{eq:metK}
\begin{split}
   G_{\alpha\beta}&=\frac{3}{2\mathcal{K}}\int_Y\omega_\alpha\wedge *_6\omega_\beta 
      =-\frac{3}{2}\left(\frac{\mathcal{K}_{\alpha\beta}}{\mathcal{K}}-\frac{3}{2}
       \frac{\mathcal{K}_\alpha\mathcal{K}_\beta}{\mathcal{K}^2}\right) \ , \\
   G_{ab}&=\frac{3}{2\mathcal{K}}\int_Y\omega_a\wedge *_6\omega_b
      =-\frac{3}{2}\frac{\mathcal{K}_{ab}}{\mathcal{K}} \ , 
\end{split}
\end{equation}
which is just the usual metric for the space of K\"ahler deformations split into odd and even part with respect to the involution $\sigma$. The inverse metrics of \eqref{eq:metK} are denoted by $G^{\alpha\beta}$ and $G^{ab}$. Similarly, for the complex structure deformations $z^{\tilde a}$ one defines the special K\"ahler metric \cite{Candelas:1990pi}
\begin{align} \label{eq:CSt}
   \mathcal{G}_{\tilde a\tilde b}\
      =\ \frac{\partial^2}{\partial  z^{\tilde a}\partial \bar z^{\tilde b}} \
      K_\text{CS}(z,\bar z) \ , &&
      K_\text{CS}(z,\bar z)\ =\ -\ln \left(-i\int_Y\Omega \wedge \bar\Omega\right) \ ,
\end{align}
which is the metric on the complex structure moduli space of the Calabi-Yau manifold~$Y$ restricted to the complex structure deformations compatible with the holomorphic involution $\sigma$ \cite{Brunner:20032004}.

The gauge kinetic matrix $\mathcal{M}_{\hat\alpha\hat\beta}$ for the bulk vector fields $V^{\hat\alpha}$ is given by \cite{SuzukiCeresole}
\begin{equation} 
   \mathcal{M}=A\inv{C}+i\inv{C} \ ,   
\end{equation}
where the matrices $A$ and $C$ are specified by the integrals
\begin{align} \label{eq:mat3}
   \bti{A}{\hat\beta}{\hat\alpha}
       =-\int_Y \beta^{\hat\alpha}\wedge *_6 \alpha_{\hat\beta} \ ,  &&
   C^{\hat\alpha\hat\beta}
       =-\int_Y \beta^{\hat\alpha}\wedge *_6 \beta^{\hat\beta} \ . 
\end{align}
Finally the quantities $\mathcal{L}_{A\bar B}$ and $\mathcal{C}_\alpha^{I\bar J}$ are related to the D7-brane fields~$\db^A$ and the D7-brane Wilson line moduli~$a_I$ respectively and are given in terms of integrals over the internal cycle~$S^\Lambda$
\begin{align} \label{eq:LC}
   \mathcal{L}_{A\bar B}=\frac{\int_{S^\Lambda}\tilde s_A\wedge\tilde s_{\bar B}}
                               {\int_Y\Omega\wedge\bar\Omega}  \ , &&
   \mathcal{C}_\alpha^{I\bar J}=\int_{S^\Lambda}\iota^*\omega_\alpha\wedge A^I\wedge\bar A^{\bar J} \ .
\end{align}

Note that this action (also without all the terms resulting from the D7-brane) has a set of global shift symmetries
\begin{align} \label{eq:shift}
   &c^a\rightarrow c^a+\theta^a \ , 
   &\rho_\alpha\rightarrow\rho_\alpha+\mathcal{K}_{\alpha bc}\dbbf^b\theta^c \ .
\end{align}
In the presence of a D7-brane wrapped on the cycle $S^\Lambda$ one of these symmetries is gauged, and therefore the action \eqref{eq:action} contains covariant derivatives for the charged fields $c^P$ and $\rho_\alpha$. These fields transform non-linearly with respect to the $U(1)$ gauge group of the D7-brane and their gauge covariant derivatives are given by
\begin{align} \label{eq:cd1}
   &\cov_\mu c^a=\partial_\mu c^a-4\kappa_4^2\mu_7\ell\delta^a_P A_\mu \ ,
   &\cov_\mu\rho_\alpha=\partial_\mu\rho_\alpha-4\kappa_4^2\mu_7
     \ell\mathcal{K}_{\alpha bP} b^b A_\mu \ .
\end{align} 


\subsection{Effective action with D7-brane fluxes} \label{sec:waction}


In the presence of D7-brane background fluxes the low energy effective action \eqref{eq:action} is modified by the fluxes $\leftup{Y}{f}$ and $\tilde f$. By performing the Kaluza-Klein reduction explicitly one obtains with $\dbbf^a=b^a-\ell\leftup{Y}{f^a}$ and $\fcharge=\ell^2\int\tilde f\wedge\tilde f$ the low energy effective action
\begin{align} \label{eq:faction}
   \mathcal{S}^{\text{E}}_f
     &=\frac{1}{2\kappa_4^2}\int \left[-R\:*_41
          +2\mathcal{G}_{\tilde a\tilde b}\dd z^{\tilde a} \wedge *_4 \dd\bar z^{\tilde b}
          +2G_{\alpha\beta}\dd v^\alpha \wedge *_4 \dd v^\beta \right. \nonumber \\
      &+\frac{1}{2}\dd(\ln \mathcal{K})\wedge *_4 \dd(\ln \mathcal{K})
          +\frac{1}{2}\dd\phi\wedge *_4 \dd\phi 
          +2e^\phi G_{ab}\dd b^a\wedge *_4 \dd b^b \nonumber \\
      &+2i\kappa_4^2\mu_7\mathcal{L}_{A\bar B}\left(e^\phi+4 G_{ab} \dbbf^a \dbbf^b
          -\frac{6v^\Lambda}{\mathcal{K}}\fcharge \right)
          \dd\db^A\wedge *_4\dd\bar\db^{\bar B} 
          +\frac{24}{\mathcal{K}}\kappa_4^2\mu_7\ell^2i\mathcal{C}^{I\bar J}_\alpha v^\alpha
          \dd a_I\wedge *_4\dd\bar a_{\bar J} \nonumber \\
      &+\frac{e^{2\phi}}{2}
          \left(\dd l+\kappa_4^2\mu_7\mathcal{L}_{A\bar B}
          \left(\dd\db^A\bar\db^{\bar B}-\dd\bar\db^{\bar B}\db^A\right)\right)\wedge  
          *_4\left(\dd l+\kappa_4^2\mu_7\mathcal{L}_{A\bar B}
          \left(\dd\db^A\bar\db^{\bar B}-\dd\bar\db^{\bar B}\db^A\right)\right) \nonumber \\
      &+2e^\phi G_{ab}
          \left(\cov c^a-l\dd b^a-\kappa_4^2\mu_7 \dbbf^a \mathcal{L}_{A\bar B}
          \left(\dd\db^A\bar\db^{\bar B}-\dd\bar\db^{\bar B}\db^A\right)\right)\wedge \nonumber \\
      &\qquad *_4 \left(\cov c^b-l\dd b^b-\kappa_4^2\mu_7 \dbbf^b 
          \mathcal{L}_{A\bar B} 
          \left(\dd\db^A\bar\db^{\bar B}-\dd\bar\db^{\bar B}\db^A\right)\right) \nonumber \\
      &+\frac{9}{2\mathcal{K}^2}G^{\alpha\beta}
          \left(\cov\rho_\alpha-\mathcal{K}_{\alpha bc}
          c^b\dd b^c-\tfrac{1}{2}\kappa_4^2\mu_7
          \left(\mathcal{K}_{\alpha bc} \dbbf^b \dbbf^c
          +\delta^\Lambda_\alpha \fcharge\right) \mathcal{L}_{A\bar B}
          \left(\dd\db^A\bar\db^{\bar B}-\dd\bar\db^{\bar B}\db^A\right)\right. \nonumber \\
      &\qquad\qquad\left. +2\kappa_4^2\mu_7\ell^2\mathcal{C}_\alpha^{I\bar J}
          \left(a_I\dd \bar a_{\bar J}-\bar a_{\bar J}\dd a_I\right)\right)\wedge \nonumber \\
      &\qquad *_4 \left(\cov\rho_\beta-\mathcal{K}_{\beta ab} c^a\dd b^b
          -\tfrac{1}{2}\kappa_4^2\mu_7
          \left(\mathcal{K}_{\beta bc}\dbbf^b \dbbf^c
          +\delta^\Lambda_\alpha \fcharge\right) \mathcal{L}_{A\bar B}
          \left(\dd\db^A\bar\db^{\bar B}-\dd\bar\db^{\bar B}\db^A\right)\right. \nonumber \\
      &\qquad\qquad\left. +2\kappa_4^2\mu_7\ell^2\mathcal{C}_\beta^{I\bar J}
          \left(a_I\dd\bar a_{\bar J}-\bar a_{\bar J}\dd a_I\right)\right) \nonumber \\
      &+\kappa_4^2\mu_7\ell^2 \left(\tfrac{1}{2}\mathcal{K}_\Lambda
          -\tfrac{1}{2}e^{-\phi}\mathcal{K}_{\Lambda ab} \dbbf^a \dbbf^b 
          -\tfrac{1}{2}e^{-\phi}\fcharge \right) F\wedge *_4F \nonumber \\
      &+\kappa_4^2\mu_7\ell^2\left(\rho_\Lambda-\mathcal{K}_{\Lambda ab}c^a \dbbf^b
          +\tfrac{1}{2}\mathcal{K}_{\Lambda ab}\dbbf^a \dbbf^bl 
          +\tfrac{1}{2}l\fcharge \right) F\wedge F \nonumber \\
      &\left.+\frac{1}{2}(\Imag \mathcal{M})_{\hat\alpha\hat\beta} 
          \dd V^{\hat\alpha}\wedge *_4 \dd V^{\hat\beta}
          +\frac{1}{2}(\Real \mathcal{M})_{\hat\alpha\hat\beta}
          \dd V^{\hat\alpha}\wedge\dd V^{\hat\beta} \right] \ , 
\end{align}
with the gauge covariant derivatives given by
\begin{equation} \label{eq:cd2}
\begin{split}
   &\cov_\mu c^a=\partial_\mu c^a-4\kappa_4^2\mu_7\ell\delta^a_P A_\mu \ , \\
   &\cov_\mu \rho_\alpha=\partial_\mu\rho_\alpha-4\kappa_4^2\mu_7
     \ell\mathcal{K}_{\alpha bP} \dbbf^b A_\mu-4\kappa_4^2\mu_7\ell Q_\alpha A_\mu \ ,
\end{split}
\end{equation}
where $Q_\alpha=\ell \int \omega_\alpha\wedge P_-\tilde f$.


\section{Spin representations and conventions} \label{sec:conv}


This appendix summarizes the conventions used in this paper. Throughout the paper pseudo Euclidean metrics have the signature $(-++\ldots)$. We use the letters of the beginning of the alphabet to denote the `flat' Lorentz frame indices whereas the letters in the middle of the alphabet are used for the `curved' coordinate indices. The sign of the epsilon symbol~$\hat\epsilon$ for the Lorentz frames is given by
\begin{align} \label{eq:epsflat}
   \hat\epsilon_{012\ldots}=
      \begin{cases} 1  & \text{Euclidean} \\ 
                   -1 & \text{Pseudo Euclidean} 
      \end{cases} \ , && 
   \hat\epsilon^{012\ldots}=+1 \ ,
\end{align} 
whereas the epsilon symbol~$\epsilon$ for the `curved' coordinates is given by
\begin{align}
   \epsilon_{012\ldots}=\det g \ , && \epsilon^{012\ldots}=1 \ ,
\end{align}
where $g$ is the metric of the Euclidean or pseudo Euclidean space. Note that $\det g$ is negative in the pseudo Euclidean case.


\subsection{Spin representations and Dirac gamma matrices} \label{sec:Spinors}


The ten-dimensional $32\times 32$ Dirac gamma matrices~$\Gamma^M$ fulfill the usual Clifford algebra
\begin{align} \label{eq:DG10}
   \ac{\Gamma^A}{\Gamma^B}=2 \eta^{AB} \ , && A,B=0,\ldots,9 \ ,
\end{align}
where $\eta^{AB}=\diag{-1,+1,\ldots,+1}$ is the metric tensor of ten-dimensional Minkowski space invariant under the Lorentz group $SO(9,1)$. Furthermore the ten-dimensional chirality matrix~$\Gamma$ is defined as
\begin{equation} \label{eq:G11}
   \Gamma=\Gamma^0\ldots\Gamma^9
     =-\frac{1}{10!}\:\epsilon_{A_0\ldots A_9}\Gamma^{A_0}\ldots\Gamma^{A_9} \ ,
\end{equation}
and which fulfills
\begin{align}
   \Gamma^2=\id \ , && \ac{\Gamma^M}{\Gamma}=0 \ .
\end{align}
The ten-dimensional Dirac spinor $\spinrep{32}$ decomposes into two Weyl representations $\spinrep{16}$ and $\spinrep{16'}$ of $SO(9,1)$ with opposite chirality, i.e. $\spinrep{16}$ is in the $+1$-eigenspace with respect to the chirality matrix \eqref{eq:G11} whereas $\spinrep{16'}$ is the $-1$-eigenspace.

In the context of compactifying the ten-dimensional space-time manifold to four dimensions, the Weyl spinors of $SO(9,1)$ must be decomposed into representations of $SO(3,1)\times SO(6)$
\begin{align} \label{eq:D10_D4}
   \spinrep{16}\rightarrow (\spinrep{2},\spinrep{4})\oplus (\spinrep{\bar 2},\spinrep{\bar 4}) \ , &&
   \spinrep{16'}\rightarrow (\spinrep{2},\spinrep{\bar 4})\oplus (\spinrep{\bar 2},\spinrep{4}) \ , 
\end{align}
where $\spinrep{2}$ and $\spinrep{\bar 2}$ are the two Weyl spinors of $SO(3,1)$ and $\spinrep{4}$ and $\spinrep{\bar 4}$ are the two Weyl spinors of $SO(6)$. In both cases these representations are complex conjugate to each other.

The ten-dimensional Dirac gamma matrices~\eqref{eq:DG10} can be given in terms of tensor products of gamma matrices $\gamma^\alpha$ and $\check\gamma^a$ of the Clifford algebras associated to the groups $SO(3,1)$ and $SO(6)$ respectively, namely
\begin{align} \label{eq:G10to4}
   \Gamma^\alpha=\hat\gamma^\alpha\otimes\id \ , &&\alpha=0,\ldots 3 \ , &&&&
   \Gamma^a=\hat\gamma\otimes\check\gamma^a \ , && a=1,\ldots,6 \ ,
\end{align}
with $\ac{\hat\gamma^\alpha}{\hat\gamma^\beta}=2\eta^{\alpha\beta}$ and $\ac{\check\gamma^a}{\check\gamma^b}=2\delta^{ab}$ and where the chirality matrices are defined as
\begin{align} \label{eq:G5}
   \hat\gamma=
       \frac{i}{4!}\epsilon_{\alpha_0\ldots\alpha_3}
       \hat\gamma^{\alpha_0}\ldots\hat\gamma^{\alpha_3} \ , &&
   \check\gamma=
       \frac{i}{6!}\epsilon_{a_1\ldots a_6}
       \check\gamma^{a_1}\ldots\check\gamma^{a_6} \ .
\end{align}
Then it is easy to check that \eqref{eq:G10to4} leads to
\begin{align}
   \ac{\Gamma^\alpha}{\Gamma^\beta}=2\eta^{\alpha\beta} \ , &&
   \ac{\Gamma^a}{\Gamma^b}=2 \delta^{ab} \ ,
\end{align}
and the ten-dimensional chirality matrix \eqref{eq:G11} becomes
\begin{equation}\label{eq:gammadecomp}
   \Gamma=\hat\gamma^5\otimes\check\gamma \ .
\end{equation}

In the analysis of the fermions arising from the space-time filling D7-branes one considers the decomposition of the ten-dimensional Weyl representations of the Lorentz group $SO(9,1)$ into representations of the subgroup~$SO(3,1)\times SO(4)\times SO(2)$. This yields
\begin{equation} \label{eq:MWdecomp}
\begin{split}
   \spinrep{16}&\rightarrow \left(\spinrep{2},\spinrep{2},\spinrep{1}\right)\oplus
   \left(\spinrep{2},\spinrep{2'},\spinrep{\bar 1}\right)\oplus
   \left(\spinrep{\bar 2},\spinrep{2},\spinrep{\bar 1}\right)\oplus
   \left(\spinrep{\bar 2},\spinrep{2'},\spinrep{1}\right) \ , \\
   \spinrep{16'}&\rightarrow \left(\spinrep{2},\spinrep{2},\spinrep{\bar 1}\right)\oplus
   \left(\spinrep{2},\spinrep{2'},\spinrep{1}\right)\oplus
   \left(\spinrep{\bar 2},\spinrep{2},\spinrep{1}\right)\oplus
   \left(\spinrep{\bar 2},\spinrep{2'},\spinrep{\bar 1}\right) \ ,
\end{split}
\end{equation}
with the two Weyl spinors $\spinrep{2}$ and $\spinrep{2'}$ of $SO(4)$ and $\spinrep{1}$ and $\spinrep{\bar 1}$ of $SO(2)$. Note that the two Weyl spinors of $SO(2)$ are again related by complex conjugation.

Similar as before the ten-dimensional Dirac gamma matrices~\eqref{eq:DG10} can be written as a tensor product of the Dirac gamma matrices $\hat\gamma^\alpha$ of $SO(3,1)$, $\gamma^a$ of $SO(4)$ and $\tilde\gamma^{\tilde a}$ of $SO(2)$
\begin{equation} \label{eq:DG}
\begin{aligned} 
   \Gamma^\alpha&=\hat\gamma^\alpha\otimes\id\otimes\id \ ,  & \alpha&=0,\ldots,3 \ , \\
   \Gamma^a&=\hat\gamma\otimes\gamma^a\otimes\id \ , & a&=1,\ldots,4 \ , \\
   \Gamma^{\tilde a}&=\hat\gamma\otimes\id\otimes\tilde\gamma^{\tilde a} \ ,
      & \tilde a&=1,2 \ ,
\end{aligned}
\end{equation}
with $\ac{\gamma^a}{\gamma^b}=2\delta^{ab}$ and $\ac{\tilde\gamma^{\tilde a}}{\tilde\gamma^{\tilde b}}=2\delta^{\tilde a\tilde b}$. The chirality matrices $\gamma$ of $SO(4)$ and $\tilde\gamma$ of $SO(2)$ are given by
\begin{align} \label{eq:G5_3}
   \gamma = -\frac{1}{4!}\epsilon_{a_1\ldots a_4}\gamma^{a_1}\ldots\hat\gamma^{a_4} \ , &&
   \tilde\gamma = -\frac{i}{2}\epsilon_{\tilde a\tilde b}\tilde\gamma^{\tilde a}
      \tilde\gamma^{\tilde b} \ .
\end{align}
As before one checks that the definition \eqref{eq:DG} gives rise to the desired anti-commutation relation
\begin{equation}
\begin{aligned}
   \ac{\Gamma^\alpha}{\Gamma^\beta}&=2\eta^{\alpha\beta} \ , & 
       \ac{\Gamma^a}{\Gamma^b}&=2 \delta^{ab} \ , &
   \ac{\Gamma^{\tilde a}}{\Gamma^{\tilde b}}&=2 \delta^{\tilde a\tilde b} \ , \\
   \ac{\Gamma^\alpha}{\Gamma^b}&=0 \ , & \ac{\Gamma^\alpha}{\Gamma^{\tilde b}}&=0 \ , &
   \ac{\Gamma^a}{\Gamma^{\tilde b}}&=0 \ ,
\end{aligned}
\end{equation}
and that the ten-dimensional chirality matrix \eqref{eq:G11} is given in terms of the lower dimensional chirality matrices \eqref{eq:G5} and \eqref{eq:G5_3}, i.e. 
\begin{equation} \label{eq:G11b}
   \Gamma=\hat\gamma\otimes\gamma\otimes\tilde\gamma \ .
\end{equation}


\newpage
\subsection*{Acknowledgments}


We would like to thank Adam Falkowski, Mariana Gra\~na, Thomas Grimm, Olaf Hohm, Wolfgang Lerche, Fernando Quevedo, Uwe Semmelmann, Stephan Stieberger, Michele Trapletti, Angel Uranga  and especially Peter Mayr for helpful discussions and correspondences. Many thanks to Stefan Pokorski and Katja Poppenh\"ager for collaborating in the early stages of this project. We are grateful to Dieter L\"ust, Peter Mayr, Susanne Reffert and Stephan Stieberger for communicating their related work \cite{Lust:2005bd} prior to publication. This work is supported by DFG -- The German Science Foundation, the European RTN Program MRTN-CT-2004-503369 and the DAAD -- The German Academic Exchange Service.

\vskip 2cm


\providecommand{\href}[2]{#2}

\begingroup
\begin{thebibliography}{99}

\bibitem{ReviewDB}
For a review see, for example,
E.~Kiritsis,
``D-branes in standard model building, gravity and cosmology,''
Fortsch.\ Phys.\  {\bf 52} (2004) 200
[arXiv:hep-th/0310001];\\
A.~M.~Uranga,
``Chiral four-dimensional string compactifications with intersecting
D-branes,''
Class.\ Quant.\ Grav.\  {\bf 20}, S373 (2003)
[arXiv:hep-th/0301032];\\
D.~L\"ust,
``Intersecting brane worlds: A path to the standard model?,''
Class.\ Quant.\ Grav.\  {\bf 21}, S1399 (2004)
[arXiv:hep-th/0401156];\\
R.~Blumenhagen,
``Recent progress in intersecting D-brane models,''
arXiv:hep-th/0412025;\\
R.~Blumenhagen, M.~Cveti\v c, P.~Langacker and G.~Shiu,
``Toward Realistic Intersecting D-Brane Models,''
arXiv:hep-th/0502005,
and references therein.

\bibitem{JP}
A.~Sagnotti,
``Open Strings And Their Symmetry Groups,''
arXiv:hep-th/0208020;\\
J.~Dai, R.~G.~Leigh and J.~Polchinski,
``New Connections Between String Theories,''
Mod.\ Phys.\ Lett.\ A {\bf 4} (1989) 2073;\\
R.~G.~Leigh,
``Dirac-Born-Infeld Action From Dirichlet Sigma Model,''
Mod.\ Phys.\ Lett.\ A {\bf 4} (1989) 2767;\\
G.~Pradisi and A.~Sagnotti,
``Open String Orbifolds,''
Phys.\ Lett.\ B {\bf 216}, 59 (1989);\\
M.~Bianchi and A.~Sagnotti,
``On The Systematics Of Open String Theories,''
Phys.\ Lett.\ B {\bf 247} (1990) 517;
``Twist Symmetry And Open String Wilson Lines,''
Nucl.\ Phys.\ B {\bf 361} (1991) 519;\\
P.~Ho\u rava,
``Strings On World Sheet Orbifolds,''
Nucl.\ Phys.\ B {\bf 327} (1989) 461;\\
E.~G.~Gimon and J.~Polchinski,
``Consistency Conditions for Orientifolds and D-Manifolds,''
Phys.\ Rev.\ D {\bf 54} (1996) 1667
[arXiv:hep-th/9601038].

\bibitem{Acharya:2002ag}
B.~Acharya, M.~Aganagic, K.~Hori and C.~Vafa,
``Orientifolds, mirror symmetry and superpotentials,''
arXiv:hep-th/0202208.

\bibitem{Brunner:20032004}
I.~Brunner and K.~Hori,
``Orientifolds and mirror symmetry,''
JHEP {\bf 0411} (2004) 005
[arXiv:hep-th/0303135];\\
I.~Brunner, K.~Hori, K.~Hosomichi and J.~Walcher,
``Orientifolds of Gepner models,''
arXiv:hep-th/0401137.

\bibitem{Grimm:2004}
T.~W.~Grimm and J.~Louis,
``The effective action of N = 1 Calabi-Yau orientifolds,''
Nucl.\ Phys.\ B {\bf 699} (2004) 387
[arXiv:hep-th/0403067];\\
T.~W.~Grimm and J.~Louis,
``The effective action of type IIA Calabi-Yau orientifolds,''
arXiv:hep-th/0412277.

\bibitem{Bachas:1995ik}
C.~Bachas,
``A Way to break supersymmetry,''
arXiv:hep-th/9503030.

\bibitem{Polchinski:1995sm}
J.~Polchinski and A.~Strominger,
``New Vacua for Type II String Theory,''
Phys.\ Lett.\ B {\bf 388}, 736 (1996)
[arXiv:hep-th/9510227].

\bibitem{Michelson:1996pn}
J.~Michelson,
``Compactifications of type IIB strings to four dimensions with  non-trivial
classical potential,''
Nucl.\ Phys.\ B {\bf 495}, 127 (1997)
[arXiv:hep-th/9610151].

\bibitem{Gukov:1999ya}
S.~Gukov, C.~Vafa and E.~Witten,
``CFT's from Calabi-Yau four-folds,''
Nucl.\ Phys.\ B {\bf 584}, 69 (2000)
[Erratum-ibid.\ B {\bf 608}, 477 (2001)]
[arXiv:hep-th/9906070].

\bibitem{Dasgupta:1999ss}
K.~Dasgupta, G.~Rajesh and S.~Sethi,
``M theory, orientifolds and G-flux,''
JHEP {\bf 9908}, 023 (1999)
[arXiv:hep-th/9908088].

\bibitem{Fluxes:19992001}
T.~R.~Taylor and C.~Vafa,
``RR flux on Calabi-Yau and partial supersymmetry breaking,''
Phys.\ Lett.\ B {\bf 474}, 130 (2000)
[arXiv:hep-th/9912152];\\
P.~Mayr,
``On supersymmetry breaking in string theory and its realization in brane
worlds,''
Nucl.\ Phys.\ B {\bf 593}, 99 (2001)
[arXiv:hep-th/0003198];\\
G.~Curio, A.~Klemm, D.~L\"ust and S.~Theisen,
``On the vacuum structure of type II string compactifications on  Calabi-Yau
spaces with H-fluxes,''
Nucl.\ Phys.\ B {\bf 609}, 3 (2001)
[arXiv:hep-th/0012213];\\
K.~Becker and M.~Becker,
``Supersymmetry breaking, M-theory and fluxes,''
JHEP {\bf 0107}, 038 (2001)
[arXiv:hep-th/0107044].

\bibitem{Giddings:2001yu}
S.~B.~Giddings, S.~Kachru and J.~Polchinski,
``Hierarchies from fluxes in string compactifications,''
Phys.\ Rev.\ D {\bf 66}, 106006 (2002)
[arXiv:hep-th/0105097].

\bibitem{Haack:2001jz}
M.~Haack and J.~Louis,
``M-theory compactified on Calabi-Yau fourfolds with background flux,''
Phys.\ Lett.\ B {\bf 507} (2001) 296
[arXiv:hep-th/0103068].

\bibitem{Becker:2002nn}
K.~Becker, M.~Becker, M.~Haack and J.~Louis,
``Supersymmetry breaking and alpha'-corrections to flux induced  potentials,''
JHEP {\bf 0206}, 060 (2002)
[arXiv:hep-th/0204254].

\bibitem{Giryavets:2003vd}
A.~Giryavets, S.~Kachru, P.~K.~Tripathy and S.~P.~Trivedi,
``Flux compactifications on Calabi-Yau threefolds,''
JHEP {\bf 0404}, 003 (2004)
[arXiv:hep-th/0312104].

\bibitem{Grana:2002}
M.~Gra\~na,
``D3-brane action in a supergravity background: The fermionic story,''
Phys.\ Rev.\ D {\bf 66} (2002) 045014
[arXiv:hep-th/0202118];\\
M.~Gra\~na,
``MSSM parameters from supergravity backgrounds,''
Phys.\ Rev.\ D {\bf 67}, 066006 (2003)
[arXiv:hep-th/0209200].

\bibitem{Kors:2003wf}
B.~K\"ors and P.~Nath,
``Effective action and soft supersymmetry breaking for intersecting D-brane
models,''
Nucl.\ Phys.\ B {\bf 681}, 77 (2004)
[arXiv:hep-th/0309167].

\bibitem{Camara:2003ku}
P.~G.~C\'amara, L.~E.~Ib\'a\~nez and A.~M.~Uranga,
``Flux-induced SUSY-breaking soft terms,''
Nucl.\ Phys.\ B {\bf 689}, 195 (2004)
[arXiv:hep-th/0311241].

\bibitem{Grana:2003ek}
M.~Gra\~na, T.~W.~Grimm, H.~Jockers and J.~Louis,
``Soft supersymmetry breaking in Calabi-Yau orientifolds with D-branes and
fluxes,''
Nucl.\ Phys.\ B {\bf 690} (2004) 21
[arXiv:hep-th/0312232].

\bibitem{Lawrence:2004zk}
A.~Lawrence and J.~McGreevy,
``Local string models of soft supersymmetry breaking,''
JHEP {\bf 0406}, 007 (2004)
[arXiv:hep-th/0401034].

\bibitem{Lust:2004}
D.~L\"ust, S.~Reffert and S.~Stieberger,
``Flux-induced soft supersymmetry breaking in chiral type IIb orientifolds
with D3/D7-branes,''
arXiv:hep-th/0406092;\\
D.~L\"ust, S.~Reffert and S.~Stieberger,
``MSSM with soft SUSY breaking terms from D7-branes with fluxes,''
arXiv:hep-th/0410074.

\bibitem{Camara:2004jj}
P.~G.~C\'amara, L.~E.~Ib\'a\~nez and A.~M.~Uranga,
``Flux-induced SUSY-breaking soft terms on D7-D3 brane systems,''
arXiv:hep-th/0408036.

\bibitem{Font:2004cx}
A.~Font and L.~E.~Ib\'a\~nez,
``SUSY-breaking Soft Terms in a MSSM Magnetized D7-brane Model,''
arXiv:hep-th/0412150.

\bibitem{Lust:2005bd}
D.~L\"ust, P.~Mayr, S.~Reffert and S.~Stieberger,
``F-theory Flux, Destabilization of Orientifolds and Soft Terms on
D7--Branes,''
arXiv:hep-th/0501139.

\bibitem{Stabilize:20022004}
S.~Kachru, M.~B.~Schulz and S.~Trivedi,
``Moduli stabilization from fluxes in a simple IIB orientifold,''
JHEP {\bf 0310}, 007 (2003)
[arXiv:hep-th/0201028];\\
R.~Blumenhagen, B.~K\"ors and D.~L\"ust,
``Moduli stabilization for intersecting brane worlds in type 0' string
theory,''
Phys.\ Lett.\ B {\bf 532}, 141 (2002)
[arXiv:hep-th/0202024];\\
R.~Blumenhagen, D.~L\"ust and T.~R.~Taylor,
``Moduli stabilization in chiral type IIB orientifold models with fluxes,''
Nucl.\ Phys.\ B {\bf 663}, 319 (2003)
[arXiv:hep-th/0303016];\\
J.~F.~G.~Cascales, M.~P.~Garc\'ia del Moral, F.~Quevedo and A.~M.~Uranga,
``Realistic D-brane models on warped throats: Fluxes, hierarchies and moduli
stabilization,''
JHEP {\bf 0402}, 031 (2004)
[arXiv:hep-th/0312051];\\
R.~D'Auria, S.~Ferrara and M.~Trigiante,
``Orientifolds, brane coordinates and special geometry,''
arXiv:hep-th/0407138.

\bibitem{Cascales:2003pt}
J.~F.~G.~Cascales and A.~M.~Uranga,
``Chiral 4d string vacua with D-branes and moduli stabilization,''
arXiv:hep-th/0311250.

\bibitem{KKLT:2003}
S.~Kachru, R.~Kallosh, A.~Linde and S.~P.~Trivedi,
``De Sitter vacua in string theory,''
Phys.\ Rev.\ D {\bf 68}, 046005 (2003)
[arXiv:hep-th/0301240];\\
S.~Kachru, R.~Kallosh, A.~Linde, J.~Maldacena, L.~McAllister and S.~P.~Trivedi,
``Towards inflation in string theory,''
JCAP {\bf 0310}, 013 (2003)
[arXiv:hep-th/0308055].

\bibitem{Denef:2004dm}
F.~Denef, M.~R.~Douglas and B.~Florea,
``Building a better racetrack,''
JHEP {\bf 0406}, 034 (2004)
[arXiv:hep-th/0404257].

\bibitem{Gorlich:2004qm}
L.~G\"orlich, S.~Kachru, P.~K.~Tripathy and S.~P.~Trivedi,
``Gaugino condensation and nonperturbative superpotentials in flux
compactifications,''
arXiv:hep-th/0407130.

\bibitem{Choi:2004sx}
K.~Choi, A.~Falkowski, H.~P.~Nilles, M.~Olechowski and S.~Pokorski,
``Stability of flux compactifications and the pattern of supersymmetry
breaking,''
JHEP {\bf 0411} (2004) 076
[arXiv:hep-th/0411066].

\bibitem{reviewcosmo}
For a review see, for example,
A.~Linde, ``Prospects of inflation,''
arXiv:hep-th/0402051;\\
V.~Balasubramanian,
``Accelerating universes and string theory,''
Class.\ Quant.\ Grav.\  {\bf 21} (2004) S1337
[arXiv:hep-th/0404075];\\
C.~P.~Burgess,
``Inflationary String Theory?,''
arXiv:hep-th/0408037,
and references therein.

\bibitem{Burgess:2003ic}
C.~P.~Burgess, R.~Kallosh and F.~Quevedo,
``de Sitter string vacua from supersymmetric D-terms,''
JHEP {\bf 0310}, 056 (2003)
[arXiv:hep-th/0309187].

\bibitem{Lust:2004cx}
D.~L\"ust, P.~Mayr, R.~Richter and S.~Stieberger,
``Scattering of gauge, matter, and moduli fields from intersecting branes,''
Nucl.\ Phys.\ B {\bf 696}, 205 (2004)
[arXiv:hep-th/0404134].

\bibitem{Jockers:2004yj}
H.~Jockers and J.~Louis,
``The effective action of D7-branes in N = 1 Calabi-Yau orientifolds,''
Nucl.\ Phys.\ B {\bf 705} (2005) 167
[arXiv:hep-th/0409098].

\bibitem{Brunner:1999jq}
I.~Brunner, M.~R.~Douglas, A.~E.~Lawrence and C.~Romelsberger,
``D-branes on the quintic,''
JHEP {\bf 0008}, 015 (2000)
[arXiv:hep-th/9906200].

\bibitem{Witten:1992fb}
E.~Witten,
``Chern-Simons gauge theory as a string theory,''
Prog.\ Math.\  {\bf 133} (1995) 637
[arXiv:hep-th/9207094].

\bibitem{Kachru:2000ih}
S.~Kachru, S.~Katz, A.~E.~Lawrence and J.~McGreevy,
``Open string instantons and superpotentials,''
Phys.\ Rev.\ D {\bf 62} (2000) 026001
[arXiv:hep-th/9912151].

\bibitem{Mayr:2001}
P.~Mayr,
``N = 1 mirror symmetry and open/closed string duality,''
Adv.\ Theor.\ Math.\ Phys.\  {\bf 5} (2002) 213
[arXiv:hep-th/0108229];\\
W.~Lerche and P.~Mayr,
``On N = 1 mirror symmetry for open type II strings,''
arXiv:hep-th/0111113;\\
W.~Lerche, P.~Mayr and N.~Warner,
``N = 1 special geometry, mixed Hodge variations and toric geometry,''
arXiv:hep-th/0208039;\\
W.~Lerche, P.~Mayr and N.~Warner,
``Holomorphic N = 1 special geometry of open-closed type II strings,''
arXiv:hep-th/0207259;\\
for a review see, W.~Lerche,
``Special geometry and mirror symmetry for open string backgrounds with N = 1
supersymmetry,''
arXiv:hep-th/0312326.

\bibitem{Balasubramanian:2004uy}
V.~Balasubramanian and P.~Berglund,
``Stringy corrections to Kaehler potentials, SUSY breaking, and the
cosmological constant problem,''
JHEP {\bf 0411}, 085 (2004)
[arXiv:hep-th/0408054].

\bibitem{Green:1996dd}
M.~B.~Green, J.~A.~Harvey and G.~W.~Moore,
``I-brane inflow and anomalous couplings on D-branes,''
Class.\ Quant.\ Grav.\  {\bf 14} (1997) 47
[arXiv:hep-th/9605033].

\bibitem{Cheung:1997az}
Y.~K.~Cheung and Z.~Yin,
``Anomalies, branes, and currents,''
Nucl.\ Phys.\ B {\bf 517}, 69 (1998)
[arXiv:hep-th/9710206].

\bibitem{Cederwall:1996}
M.~Cederwall, A.~von Gussich, B.~E.~W.~Nilsson and A.~Westerberg,
``The Dirichlet super-three-brane in ten-dimensional type IIB  supergravity,''
Nucl.\ Phys.\ B {\bf 490} (1997) 163
[arXiv:hep-th/9610148];\\
M.~Cederwall, A.~von Gussich, B.~E.~W.~Nilsson, P.~Sundell and A.~Westerberg,
``The Dirichlet super-p-branes in ten-dimensional type IIA and IIB
supergravity,''
Nucl.\ Phys.\ B {\bf 490}, 179 (1997)
[arXiv:hep-th/9611159].

\bibitem{Bergshoeff:1996tu}
E.~Bergshoeff and P.~K.~Townsend,
``Super D-branes,''
Nucl.\ Phys.\ B {\bf 490}, 145 (1997)
[arXiv:hep-th/9611173].

\bibitem{DeWolfeDeAlwisBuchel}
O.~DeWolfe and S.~B.~Giddings,
``Scales and hierarchies in warped compactifications and brane worlds,''
Phys.\ Rev.\ D {\bf 67}, 066008 (2003)
[arXiv:hep-th/0208123];\\
S.~P.~de Alwis,
``On potentials from fluxes,''
Phys.\ Rev.\ D {\bf 68} (2003) 126001
[arXiv:hep-th/0307084];\\
S.~P.~de Alwis,
``Brane worlds in 5D and warped compactifications in IIB,''
arXiv:hep-th/0407126;\\
A.~Buchel,
``On effective action of string theory flux compactifications,''
Phys.\ Rev.\ D {\bf 69}, 106004 (2004)
[arXiv:hep-th/0312076].

\bibitem{Vafa:1995gm}
C.~Vafa and E.~Witten,
``Dual string pairs with N = 1 and N = 2 supersymmetry in four  dimensions,''
Nucl.\ Phys.\ Proc.\ Suppl.\  {\bf 46}, 225 (1996)
[arXiv:hep-th/9507050].

\bibitem{Dabholkar:1996pc}
A.~Dabholkar and J.~Park,
``Strings on Orientifolds,''
Nucl.\ Phys.\ B {\bf 477} (1996) 701
[arXiv:hep-th/9604178].

\bibitem{Candelas:1990pi}
P.~Candelas and X.~de la Ossa,
``Moduli Space Of Calabi-Yau Manifolds,''
Nucl.\ Phys.\ B {\bf 355} (1991) 455.

\bibitem{StefanskiScrucca}
B.~J.~Stefanski,
``Gravitational couplings of D-branes and O-planes,''
Nucl.\ Phys.\ B {\bf 548} (1999) 275
[arXiv:hep-th/9812088];\\
C.~A.~Scrucca and M.~Serone,
``Anomalies and inflow on D-branes and O-planes,''
Nucl.\ Phys.\ B {\bf 556} (1999) 197
[arXiv:hep-th/9903145].

\bibitem{Blumenhagen:2002wn}
R.~Blumenhagen, V.~Braun, B.~K\"ors and D.~L\"ust,
``Orientifolds of K3 and Calabi-Yau manifolds with intersecting D-branes,''
JHEP {\bf 0207} (2002) 026
[arXiv:hep-th/0206038].

\bibitem{Cremmer:1982en}
E.~Cremmer, S.~Ferrara, L.~Girardello and A.~Van Proeyen,
``Yang-Mills Theories With Local Supersymmetry: Lagrangian, Transformation
Laws And Superhiggs Effect,''
Nucl.\ Phys.\ B {\bf 212} (1983) 413.

\bibitem{Wess:1992}
J.~Wess and J.~Bagger,
``Supersymmetry And Supergravity,''
Princeton University Press, Princeton, 1992.

\bibitem{Haack:1999zv}
M.~Haack and J.~Louis,
``Duality in heterotic vacua with four supercharges,''
Nucl.\ Phys.\ B {\bf 575} (2000) 107
[arXiv:hep-th/9912181].

\bibitem{Hsu:2003cy}
J.~P.~Hsu, R.~Kallosh and S.~Prokushkin,
``On brane inflation with volume stabilization,''
JCAP {\bf 0312} (2003) 009
[arXiv:hep-th/0311077].

\bibitem{Berg:2004ek}
M.~Berg, M.~Haack and B.~K\"ors,
``Loop corrections to volume moduli and inflation in string theory,''
arXiv:hep-th/0404087.

\bibitem{Dudas:2000ff}
E.~Dudas and J.~Mourad,
``Brane solutions in strings with broken supersymmetry and dilaton
tadpoles,''
Phys.\ Lett.\ B {\bf 486} (2000) 172
[arXiv:hep-th/0004165].

\bibitem{Blumenhagen:2001te}
R.~Blumenhagen, B.~K\"ors, D.~L\"ust and T.~Ott,
``The standard model from stable intersecting brane world orbifolds,''
Nucl.\ Phys.\ B {\bf 616} (2001) 3
[arXiv:hep-th/0107138].

\bibitem{Fischler:1986}
W.~Fischler and L.~Susskind,
``Dilaton Tadpoles, String Condensates And Scale Invariance,''
Phys.\ Lett.\ B {\bf 171} (1986) 383;\\
W.~Fischler and L.~Susskind,
``Dilaton Tadpoles, String Condensates And Scale Invariance. 2,''
Phys.\ Lett.\ B {\bf 173} (1986) 262.

\bibitem{Marino:1999af}
M.~Mari\~no, R.~Minasian, G.~W.~Moore and A.~Strominger,
``Nonlinear instantons from supersymmetric p-branes,''
JHEP {\bf 0001} (2000) 005
[arXiv:hep-th/9911206].

\bibitem{Blumenhagen:2000wh}
R.~Blumenhagen, L.~G\"orlich, B.~K\"ors and D.~L\"ust,
``Noncommutative compactifications of type I strings on tori with  magnetic
background flux,''
JHEP {\bf 0010} (2000) 006
[arXiv:hep-th/0007024].

\bibitem{Cvetic:2001nr}
M.~Cveti\v c, G.~Shiu and A.~M.~Uranga,
``Chiral four-dimensional N = 1 supersymmetric type IIA orientifolds from
intersecting D6-branes,''
Nucl.\ Phys.\ B {\bf 615}, 3 (2001)
[arXiv:hep-th/0107166].

\bibitem{Howe:1983sr}
P.~S.~Howe and P.~C.~West,
``The Complete N=2, D = 10 Supergravity,''
Nucl.\ Phys.\ B {\bf 238} (1984) 181.

\bibitem{Bergshoeff:19972000}
E.~Bergshoeff, R.~Kallosh, T.~Ortin and G.~Papadopoulos,
``kappa-symmetry, supersymmetry and intersecting branes,''
Nucl.\ Phys.\ B {\bf 502} (1997) 149
[arXiv:hep-th/9705040];\\
E.~A.~Bergshoeff, M.~de Roo and A.~Sevrin,
``Non-Abelian Born-Infeld and kappa-symmetry,''
J.\ Math.\ Phys.\  {\bf 42} (2001) 2872
[arXiv:hep-th/0011018].

\bibitem{Grisaru:1997ub}
M.~T.~Grisaru, M.~E.~Knutt-Wehlau and W.~Siegel,
``A superspace normal coordinate derivation of the density formula,''
Nucl.\ Phys.\ B {\bf 523} (1998) 663
[arXiv:hep-th/9711120].

\bibitem{Marolf:2003vf}
D.~Marolf, L.~Martucci and P.~J.~Silva,
``Actions and fermionic symmetries for D-branes in bosonic backgrounds,''
JHEP {\bf 0307}, 019 (2003)
[arXiv:hep-th/0306066].

\bibitem{Millar:2000ib}
K.~Millar, W.~Taylor and M.~Van Raamsdonk,
``D-particle polarizations with multipole moments of higher-dimensional
branes,''
arXiv:hep-th/0007157.

\bibitem{Bergshoeff:1999bx}
E.~Bergshoeff, M.~de Roo, B.~Janssen and T.~Ortin,
``The super D9-brane and its truncations,''
Nucl.\ Phys.\ B {\bf 550} (1999) 289
[arXiv:hep-th/9901055].

\bibitem{Sen:1996vd}
A.~Sen,
``F-theory and Orientifolds,''
Nucl.\ Phys.\ B {\bf 475}, 562 (1996)
[arXiv:hep-th/9605150].

\bibitem{Witten:1991zz}
E.~Witten,
``Mirror manifolds and topological field theory,''
arXiv:hep-th/9112056.

\bibitem{Hori:2000ck}
K.~Hori, A.~Iqbal and C.~Vafa,
``D-branes and mirror symmetry,''
arXiv:hep-th/0005247.

\bibitem{Bershadsky:1993cx}
M.~Bershadsky, S.~Cecotti, H.~Ooguri and C.~Vafa,
``Kodaira-Spencer theory of gravity and exact results for quantum string
amplitudes,''
Commun.\ Math.\ Phys.\  {\bf 165}, 311 (1994)
[arXiv:hep-th/9309140].

\bibitem{Strominger:1985ks}
A.~Strominger,
``Yukawa Couplings In Superstring Compactification,''
Phys.\ Rev.\ Lett.\  {\bf 55} (1985) 2547.

\bibitem{SuzukiCeresole}
H.~Suzuki,
``Calabi-Yau compactification of type IIB string and a mass formula of the
extreme black holes,''
Mod.\ Phys.\ Lett.\ A {\bf 11} (1996) 623
[arXiv:hep-th/9508001];\\
A.~Ceresole, R.~D'Auria and S.~Ferrara,
``The Symplectic Structure of N=2 Supergravity and its Central Extension,''
Nucl.\ Phys.\ Proc.\ Suppl.\  {\bf 46}, 67 (1996)
[arXiv:hep-th/9509160].

\end{thebibliography}
\endgroup
\end{document}